\documentclass[10pt,journal,compsoc]{IEEEtran}

\ifCLASSOPTIONcompsoc
  \usepackage[nocompress]{cite}
\else
  \usepackage{cite}
\fi

\ifCLASSINFOpdf
\else
\fi

\usepackage{hyperref}

\usepackage{enumitem}
\setlist{nolistsep}

\usepackage{setspace}
\usepackage{textcomp}
\usepackage{color,soul}
\usepackage{url}
\usepackage[disable]{todonotes}
\usepackage{caption}

\usepackage[final, todonotes={disable}]{changes}

\begin{document}
\title{Tilt Map: Interactive Transitions Between \\ Choropleth Map, Prism Map and Bar Chart \\ in Immersive Environments}

\author{
  Yalong Yang, Tim Dwyer, Kim Marriott, Bernhard Jenny and Sarah Goodwin%
\IEEEcompsocitemizethanks{\IEEEcompsocthanksitem Yalong Yang was with the Department of Human-Centred Computing, Faculty of Information Technology, Monash University, Australia. He is now with School of Engineering and Applied Sciences, Harvard University, Cambridge, MA, 02138. \protect E-mail: yalongyang@g.harvard.edu

\IEEEcompsocthanksitem Tim Dwyer, Kim Marriott, Bernhard Jenny and Sarah Goodwin are with the Department of Human-Centred Computing, Faculty of Information Technology, Monash University, Australia. \protect 
E-mail: \{tim.dwyer, kim.marriott, bernie.jenny, sarah.goodwin\}@ monash.edu
}%
\thanks{Manuscript received April 19, 2005; revised August 26, 2015.}}

\markboth{Journal of \LaTeX\ Class Files,~Vol.~14, No.~8, August~2015}%
{Shell \MakeLowercase{\textit{et al.}}: Bare Demo of IEEEtran.cls for Computer Society Journals}

\IEEEtitleabstractindextext{%
\begin{abstract}
We introduce \emph{Tilt Map}, a novel interaction technique for intuitively transitioning between 2D and 3D map visualisations in immersive environments.  Our focus is visualising data associated with areal features on maps, for example, population density by state.
\emph{Tilt Map} transitions from 2D choropleth maps to 3D prism maps to 2D bar charts to overcome the limitations of each.  Our paper includes two user  studies.  
The first study compares subjects' task performance interpreting population density data using 2D choropleth maps and 3D prism maps in virtual reality (VR).  We observed greater task accuracy with prism maps, but faster response times with choropleth maps. The complementarity of these views inspired our hybrid \emph{Tilt Map} design.
Our second study compares \emph{Tilt Map} to: a side-by-side arrangement of the various views; and interactive toggling between views. The results indicate benefits for \emph{Tilt Map} in user preference; and accuracy (versus side-by-side) and time (versus toggle). 
\end{abstract}

\begin{IEEEkeywords}
Immersive analytics, Mixed / augmented reality, Virtual reality, Geographic visualization, Interaction techniques.
\end{IEEEkeywords}}

\maketitle

\IEEEdisplaynontitleabstractindextext

\IEEEpeerreviewmaketitle

\begin{figure*}
  \centering
  \includegraphics[width=.86\textwidth]{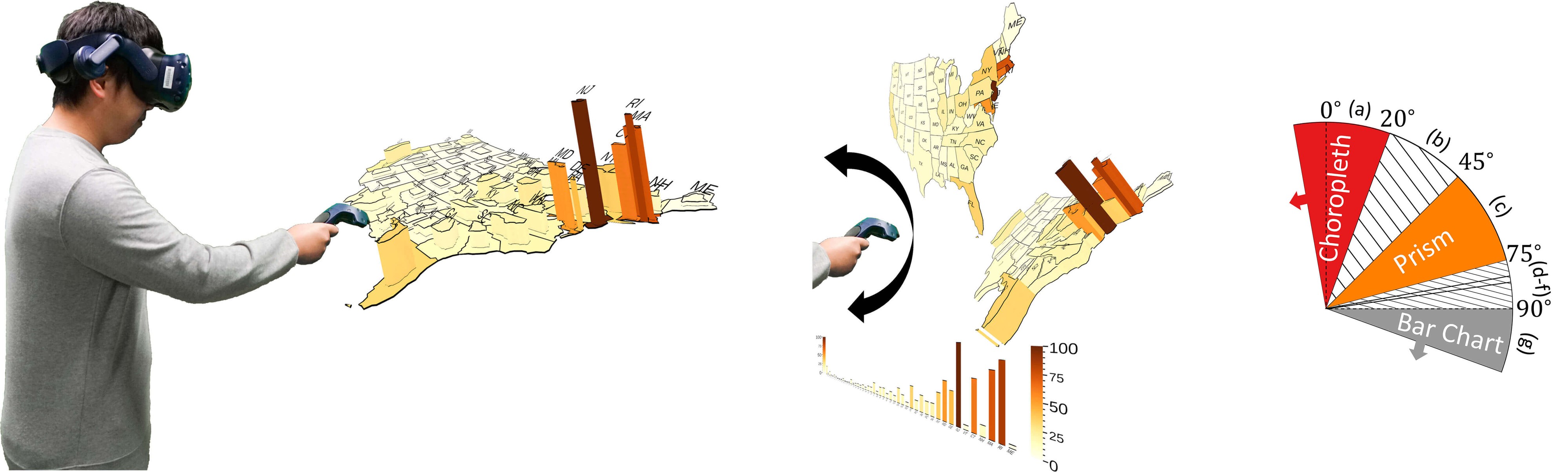}  
  \captionof{figure}{Orientation-dependent visualisation with a \textit{Tilt Map:} The user can tilt a prism map (left) to morph it to either a vertical choropleth map or a bar chart (middle). The transition stages with angle intervals were labeled from \emph{a} to \emph{g} (right). We demonstrate the transition stages with visualisations in Fig.~\ref{fig:Exp-02-magic}. \label{fig:teaser}}
  \vspace{-0.5em}
\end{figure*}

\IEEEraisesectionheading{\section{Introduction}\label{sec:introduction}}

\IEEEPARstart{C}{horopleth} maps are arguably the most widely used visualisation for showing data linked to geographic areas, such as population density ~\cite{dent:2008, slocum2009thematic}. These maps colour or shade the areas on the map to indicate the associated values. A less common way of showing such area-linked data is the \emph{prism map}~\cite{kraak2011cartography, field2018cartography} where areas are extruded into the third dimension so that the height of the ``prism'' represents the associated value.
\added{With the arrival of commodity head-mounted displays (HMDs) for virtual reality (VR) and augmented reality (AR), we can expect to see choropleth maps, prism maps, and other area-linked geographic visualisations used in immersive applications. However, area-linked geographic visualization is under explored in immersive environments and the trade-off between the two-dimensional choropleth map and the three-dimensional prism map
is currently unclear.}

In this research we explore whether traditional 2D choropleth maps are the best way to show area-linked data in such immersive environments, or whether other visualisations, such as a prism maps that make use of a third dimension, or some combination of these may be better. 
This paper makes three main contributions.

\noindent\textbf{The first contribution} is a controlled study comparing choropleth, prism and 
coloured prism maps for the first time in VR.
We found that participants were more accurate using the prism maps but faster using the choropleth maps. 
Participants preferred the coloured prism map, but raised some concern about occlusion.  
These results accord with previous studies comparing choropleth and prism maps in non-VR settings, confirming a trade-off between faster choropleth maps and more accurate prism maps.

\noindent\textbf{Our second contribution} 
is \emph{Tilt Map}, a novel interactively-controlled transition between three views: a choropleth, a prism map and a bar chart. We incorporate a bar chart as an additional view, as we thought this would aid comparison tasks and alleviate the difficulties of 3D occlusion and perspective foreshortening.
In the \emph{Tilt Map}, orientation determines the view. With a vertical orientation, the choropleth map is shown. As the viewer tilts the map it morphs into the prism map, and when the map nears a horizontal orientation, it morphs into the bar chart (Fig.~\ref{fig:teaser}).
We believe this use of orientation to select the view that is appropriate to the user's view angle is new and opens up many possibilities for the design of immersive visualisations.
    
\noindent\textbf{Our third contribution} is a controlled study comparing \emph{Tilt Map} with  two other conditions:  a \emph{Side-by-Side} complementary arrangement (of choropleth, coloured prism, and bar chart) and a \emph{Toggle} representation, which switches between the three views with a controller click.
We evaluated user preference and performance. 
The results indicate benefits for \emph{Tilt Map} in user preference and accuracy (versus side-by-side) and time (versus toggle).

Our work contributes to the growing body of knowledge of how to present data with a geographical embedding in immersive environments~\cite{yang2018maps,yang2019origin,Wagner2019}. It provides further evidence that for geographically embedded data, there can be benefits in utilising a third dimension to show the data variable. 
We also introduce a new kind of interaction specifically suited to immersive data visualisation. Just as large tiled wall displays allow the use of proxemic interaction to naturally control the choice of presentation~\cite{jakobsen2013information}, the ability to hold and tilt virtual artefacts, such as maps, in immersive environments provides an intuitive embodied method for transitioning between different views, with the aim of providing the view that is best suited to the viewing angle.

\vspace{-0.7em}
\section{Related Work}

The choropleth map---where area-based values are encoded via colour, shading or pattern---is one of the most commonly used thematic map types~\cite{dent:2008, slocum2009thematic, richardson_choropleth_2017}.
\deleted{
Whilst mapping the information geographically is fundamental to reveal spatial patterns, the attribute values are abstracted to brightness. }
\added{ 
Whilst using the positions in the display space for representing geographic information is fundamental to reveal spatial patterns, the attribute values have to be represented by non-spatial visual variables, such as colour and brightness. 
Raw-total values are avoided on choropleth maps because large geographic areas likely have large corresponding mapped quantities.
Choropleth maps with non-uniform areas should instead display data in the form of density measures, proportions, or ratios~\cite{dent:2008, slocum2009thematic, field2018cartography}.}

A prism map 
is a 3D choropleth map with extruded height to encode a numerical attribute~\cite{kraak2011cartography, field2018cartography}. Prism maps are quite uncommon. Most cartographic textbooks mention them only \textit{en passant} when discussing choropleth maps, with the recent exception of Field, who mentions that \textit{``prism maps are predominantly used for visual impact''}, and speculates that the unfamiliarity with prism maps hampers their understanding~\cite{field2018cartography}.
\added{Although less common, cartographers have used prism maps for many years, especially since computer software for the generation and animation of prism maps became available~\cite{tobler:1973, franklin1978, moellering1980real, Nimmer:1982uu}.
For static prism maps, it has been shown that readers consistently associate prism height, and not prism volume, with data value~\cite{cuff1979ratios}. This is the case for both absolute values and proportional values~\cite{cuff1979ratios}, which implies that converting quantities to densities or ratios is not necessary for prism maps.}

The question of whether it is preferable to use choropleth maps or prism maps has no immediate answer. Height (the visual variable used by prism maps) is far more accurate for interpreting associated quantitative values than brightness (the visual variable commonly used by choropleth maps)~\cite{cleveland1984graphical}, but perspective foreshortening and oblique viewing angles can result in  distortion and excessive occlusion in prism maps. Few studies compare prism maps to choropleth or other thematic map types. One study~\cite{Niedomysl:2013ko} evaluated short-term learning benefits of choropleth maps and prism maps, using uniform gridded areas for both map types instead of the more common irregular areas. 
It found  that the choropleth map improved participants' ability to correctly assess detailed information (ranking of city populations), while the prism map appeared slightly more useful for reading general patterns (overall population distribution). Another study compared prism maps and area-proportional circle maps, and found similar reading accuracy for both~\cite{cuff1979ratios}. Popelka~\cite{Popelka:2018cy} compared prism maps, standard choropleth maps and illuminated choropleth maps (where illumination effects give flat areas a subtle 3D prism-like appearance~\cite{Stewart:2010hi}). The task involved comparing the values of two marked areas. Prism maps resulted in more accurate but slower responses than choropleth maps. 
Meanwhile, studies by Bleisch, Dykes \& Nebiker~\cite{Bleisch2008} and Seipel \& Carvalho~\cite{Seipel2012} indicate that reading accuracy of bar charts heights in 2D and 3D maps is similar. Creating 3D maps is also supported in many commercial products such as ArcGIS, Mapbox, kepler.gl, rayshader. Therefore, although 3D mapping techniques are not as popular as 2D ones, it seems reasonable to assume that encoding quantitative values with prism heights is a practical design choice.

The mentioned studies have used static maps on paper or standard 2D computer displays. 
Our research provides the first comparison of choropleth and prism maps in an immersive environment where the viewing angle can be easily adjusted.
In general, there has been surprisingly little research into thematic cartography in immersive environments.
Only recently have researchers begun to systematically investigate immersive geospatial data visualisation~\cite{Marriott2018}. 
Yang \textit{et al.} explored immersive visualisation of origin-destination flow maps~\cite{yang2019origin} and of maps and globes in virtual reality~\cite{yang2018maps,phdthesis}. They found clear benefits for the use of 3D representations. 
\added{Quang and Jenny placed bar graphics in a virtual landscape and found that linking the bars with bar charts and maps with bars improves performance~\cite{quach2020}.}
Wagner \textit{et al.} compared space-time cubes in virtual reality and on 2D displays~\cite{Wagner2019}. They found similar tasks performance but the immersive version received higher subjective usability scores.
Furthermore, immersive environments support embodied interaction~\cite{cordeil2017imaxes} and allow the map reader to adjust position, size and scale in engaging ways~\cite{satriadiaugmented}. 
Our research combines embodied interaction, animated transitions between data graphics~\cite{heer2007animated} and interactive 3D geovisualisation~\cite{herman2018evaluation}, resulting in a novel type of cartographic visualisation that adjusts to the tilt angle.

\begin{figure*}[!t]
	\centering
	\includegraphics[width=.98\textwidth]{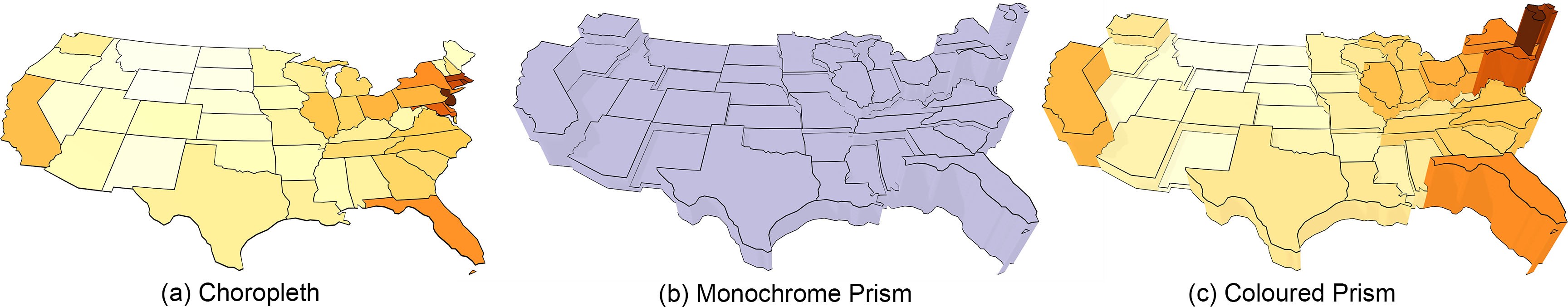}
	\caption{Study 1: Evaluated \emph{Choropleth}, \emph{Monochrome Prism} and \emph{Coloured Prism} maps, demonstrating US examples.}
	\label{fig:Exp-01-Vis}
\end{figure*}

\section{Study 1: Prism Map vs Choropleth Map}
\label{sec:study-one}
Our first user study compared three different visual representations of areal population density data in VR: 2D \emph{Choropleth} map;
3D \emph{Monochrome Prism} map;
and 3D double-encoded \emph{Coloured Prism} map (Fig.~\ref{fig:Exp-01-Vis}). 
A previous study~\cite{Popelka:2018cy} using 2D desktop displays  found that task performance with prism maps was more accurate but slower than with choropleth maps. 
We wished to see if this changed with the additional depth cues provided by head-tracking and stereoscopic presentation in modern VR environments.
\added{We evaluated the visualization conditions with one elementary task and one synoptic task (see Section~\ref{sec:exp-1}).}

\subsection{Visualisations and Interactions}
\label{sec:exp-01-vis}
\added{We map areal population density data in different encodings in our tested conditions.}
\vspace{0.5em}

\noindent\textbf{\emph{Monochrome Prism:}} We mapped the density values linearly to height.
\vspace{0.5em}

\noindent\textbf{\emph{Choropleth:}} In order to compare directly to height, we used a continuous sequential colour scheme for density values, resulting in an unclassed choropleth map. The colour scheme was derived through linear interpolation of the \emph{YlOrBr} palette from ColorBrewer~\cite{Harrower:2003jm}. This colour scheme was chosen as it is colour-blind friendly~\cite{Harrower:2003jm} and popular in cartography literature (e.g.\ in~\cite{Beecham:2016hd,Yang:2017cy}).
\vspace{0.5em}

\noindent\textbf{\emph{Coloured Prism:}} We double-encoded the density values linearly with height (as in \emph{Monochrome Prism}) and with colour (as in \emph{Choropleth}).
\vspace{0.5em}

\noindent\textbf{Encodings common to all conditions:} \textbf{(a)} Borders were added to the boundaries of geographic areas as pilot tests revealed participants preferred them; \textbf{(b)} soft shadow was enabled in both prism map conditions, but was not necessary for the 2D \emph{Choropleth} maps; \textbf{(c)} legends were placed at top, bottom, left, and right sides of all maps. Legends ranged from 0 to 100 with ticks every 5 and labelled ticks at 0, 25, 50, 75 and 100. For \emph{Choropleth}, the legends lay in the plane of the maps to show colour mapping. For \emph{Monochrome Prism}, the legends were cylinders standing perpendicularly on the plane of the base map and extruded in parallel to the prism heights. \emph{Coloured Prism} legends were the same as \emph{Monochrome Prism} but also coloured as for \emph{Choropleth}.

\noindent\textbf{Interaction:} We provided the same interactions for all three visualisations. First, viewers could move in space to change their viewpoint. Second, viewers could pick up the map using a standard hand-held VR controller, and reposition or rotate it in 3D space using clutched grasping with the controller trigger.
Users could not resize or scale the visualisations but could physically zoom by moving closer.
We did not allow other interactions such as filtering as we wished to focus on base-line readability of the representations.

\subsection{Experiment}
\label{sec:exp-1} 

\added{In this subsection, we first introduce the \emph{tasks} of the user study and the way we create task \emph{data}. We then report details of the user study including: \emph{experimental set-up}, \emph{design and procedure}, \emph{participants} and \emph{measures}.}

\vspace{0.5em}
\noindent\textbf{Tasks:} 
\todo[inline]{Reveiwer: choropleth maps are not designed for quantitative tasks, should be ordinal. Height has been proved to be more accurate than color.}
\todo[inline,color=green!40]{We: tasks were built on top of previous studies.}
Researchers~\cite{Andrienko:2006ek,Roth:2013ja,Yang:2019ww} distinguish tasks in geographic data visualisation into two levels: \emph{elementary} and \emph{synoptic}. Elementary tasks refer to single elements while synoptic tasks involve a set of elements. Following this taxonomy and previous related studies~\cite{Niedomysl:2013ko,Popelka:2018cy,Stewart:2010hi}, we designed one task of each type for the first study:

\noindent\emph{\textbf{Area-Comparison Task:} Compare the density values of two given geographic areas.} 
Initially, in our pilot study, we asked participants to identify the area with the larger density value in two given geographic areas. 
The same task was tested by~\cite{Popelka:2018cy,Stewart:2010hi} on 2D computer displays.
We found participants can answer this question easily with very high accuracy in all conditions.
Inspired by~\cite{Jansen:2015bk,Spence:2016fk}, instead of a binary result, we asked the participants to perform the more difficult task of estimating the numeric difference between two given geographic areas.

\noindent\emph{\textbf{Region Task:} Estimate the population density of a region consisting of contiguous marked areas on the map.} 
A similar task was tested by Niedomysl \textit{et al.}~\cite{Niedomysl:2013ko} on printed A4 size maps.
We randomly chose regions of 5 contiguous states in each US map and 15--20 contiguous LADs in each UK map.
The correct answer is the area-weighted average of population density across geographic areas of the region.
We explained this task in detail with examples to make sure participants fully understood, emphasising that larger geographic areas contribute more to the total region weight. After the explanation all participants reported they understood. Again, participants needed to provide a numeric answer. 

In pilots, we highlighted the borders of target areas (i.e.\ the two given areas in the area-comparison task and the set of contiguous areas in the region task), following Harrower~\cite{harrower:2007}.
However, participants reported extra effort to visually search for the two targets in the area-comparison task. As the intention of this study was not to examine the time for visual search, we chose to further mark the two areas with leader lines, which proved adequate for rapid target identification. 
In region tasks, no participant reported difficulty in identifying the target regions as the highlighted borders of multiple adjacent geographic areas made the region highly visually salient. Thus, we did not use leader lines in region tasks.

For the area-comparison task, we anticipated that distance between the two targets was likely to affect the performance. We randomly sampled pairs of areas and computed the great-circle distance between areas. We then created two categories for the area-comparison tasks: \emph{close} and \emph{far}. We considered great-circle distances below 3\textdegree\ in the US and below 0.5\textdegree\ in the UK as \emph{close} and within 25\textdegree--28\textdegree\ in the US and within 5\textdegree--5.5\textdegree\ in the UK as \emph{far}. For the region task, we identified that the degree of variation of density in the region can affect performance. For example, in the extreme case there is no variation, and the participant only needs to ascertain the value of a single geographic area to answer the region task. We used the coefficient of variation (CV)~\cite{Brown:1998dj} to measure the variation of density values. Again, we sampled CVs from random regions. Based on the distribution, we decided to use CVs within 40\%--60\% for all region tasks.

We mapped the data values linearly to colour and height, indicated by legends as described in Sec.~\ref{sec:exp-01-vis} in the range 0--100. Piloting revealed 
that results below 20 and above 80 were too easy to distinguish. Thus, we controlled the results of tasks to the range of 20--40, 40--60 and 60--80.

\vspace{0.5em}
\noindent\textbf{Data:} 
We used 2018 population density data for the United States (US) \cite{usdata} and the United Kingdom (UK) \cite{ukdata}. For the US, we used the conterminous 48 States and
for the UK, we used the 391 Local Authority Districts (LAD). 
\added{The original datasets had very skewed distributions, as they contained a small number of areas with high population density. This would have made it trivial for participants to provide correct answers. We therefore applied a square root transformation to make the tasks more challenging and to ensure the map visualisations used the entire range of colour or height variation. 
}

\added{To minimise the learning effect,} we used different data for each question. We generated data based on the transformed population density data. As spatial autocorrelation structure is important in spatial analysis, we used the  technique from~\cite{Beecham:2016hd} to ensure the Moran's \textit{I} of the generated data was the same as that of the original data.

\vspace{0.5em}
\noindent\textbf{Experimental Set-up:}
We used an HTC Vive Pro with 110\textdegree~ field of view, 2160$\times$1200 pixels resolution and 90Hz refresh rate. The PC was equipped with an Intel i7-6700K 4.0GHz processor and NVIDIA GeForce GTX 1080 graphics card. Only one hand-held VR controller was needed in the experiment: participants could use this to reposition and rotate the map in 3D space. The frame rate was around 110~FPS.

Visuals were positioned comfortably within users' reach. The map was created on top of a transparent quadrilateral measuring 1$\times$1m, and placed 0.6m in front of the participants' eye position and 0.1m below it, tilted to 45\textdegree. We repositioned the map at the beginning of every question. The height of the \emph{Monochrome Prism} and \emph{Coloured Prism} maps was linearly mapped to the range of 0$-$20~cm.

\vspace{0.5em}
\noindent\textbf{Design and Procedure:}
The experiment was within-subjects: 3 visualisations $\times$ 2 task $\times$ 2 datasets $\times$ 3 answer ranges $\times$ 2 repetitions (1 close and 1 far for area-comparison task) = 72 responses per participant and lasted 1.5 hours on average. 
\added{We used the Latin square design to balance the order of the three tested visualisations, i.e. each visualisation occurred in every position in the ordering for an equal number of participants.}

Participants were first given a brief introduction to the experiment. After putting on the VR headset, we asked them to adjust it so as to clearly see the sample text in front of them. Two types of training were provided: interaction training and task training. 
\emph{1) Interaction training} was conducted when each visualisation was presented  for the first time. The participants were introduced to the visualisation and encoding. Then we gave them sufficient time to familiarise themselves with the VR headset, controller, visualisation, encoding, and repositioning and rotating interactions.
\emph{2) Task training} was conducted when each condition 
(task $\times$ visualisation) was presented for the first time. Two sample tasks, differing from the experimental tasks, were given to participants with unlimited time. After participants finished a training task, we displayed the correct answer and allowed them to confirm the answer with the visualisation. We asked the participants to check their strategies both when they were doing the training tasks and after the correct answers were shown.

Participants first completed the area-comparison task, then the region task with a 5-minute break in between. For each task,  the three visualisations  were presented in a counterbalanced order. After completing the tasks, a post-hoc questionnaire recorded feedback on: (1) preference ranking of visualisations in terms of visual design and ease of use for the tasks; (2) rated confidence with each visualisation with a five-point Likert scale; (3) advantages and disadvantages of each visualisation; (4) strategies for different visualisations; and (5) background information about the participant. The questionnaire listed visualisations in the same order as presented in the experiment.

\vspace{0.5em}
\noindent\textbf{Participants:}
We recruited 12 participants (2 female and 10 male) from our university. All had normal or corrected-to-normal vision and included university students and researchers. All participants were within the age group 20--30. VR experience varied: 7 participants had less than 5 hours of prior VR experience, 2 had 6--20~h, and 3 had more than 20~h.  
While we used colour encoding in the study, we chose a colour-blind safe colour scheme (see Sec.~\ref{sec:exp-01-vis}). Therefore, we did not test participants for colour blindness. 

\vspace{0.5em}
\noindent\textbf{Measures:}
We measured time from first render of the visualisation to double-click of controller trigger. After double-click, the visualisation was replaced by a slider to select an integer answer from 0--100. Participants could 
precisely step adjust their answer by tapping the sides of the touchpad.
We report error rate as absolute difference between participants' answers and the correct answers.
We also recorded the position and rotation of the headset, controller and map every 0.1~seconds.

\begin{figure}
	\centering 
	\includegraphics[width=\columnwidth]{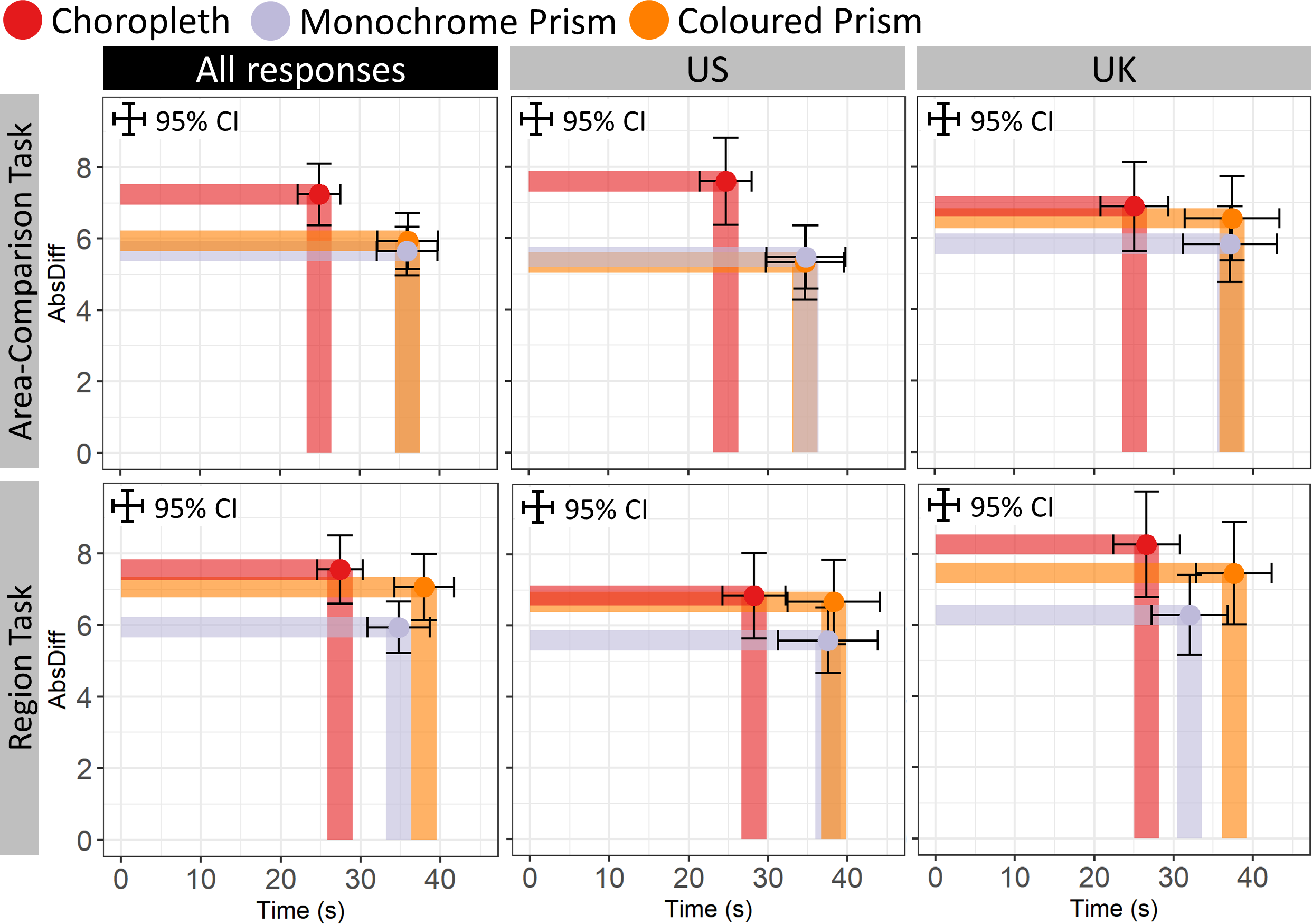}
	\caption{Study 1 --- Accuracy and response time with 95\% confidence intervals (\textit{AbsDiff} = absolute difference between participants' answers and the correct answers.)}
	\label{fig:Exp-01-results}
\end{figure}
\begin{figure}
	\centering 
	\includegraphics[width=0.9\columnwidth]{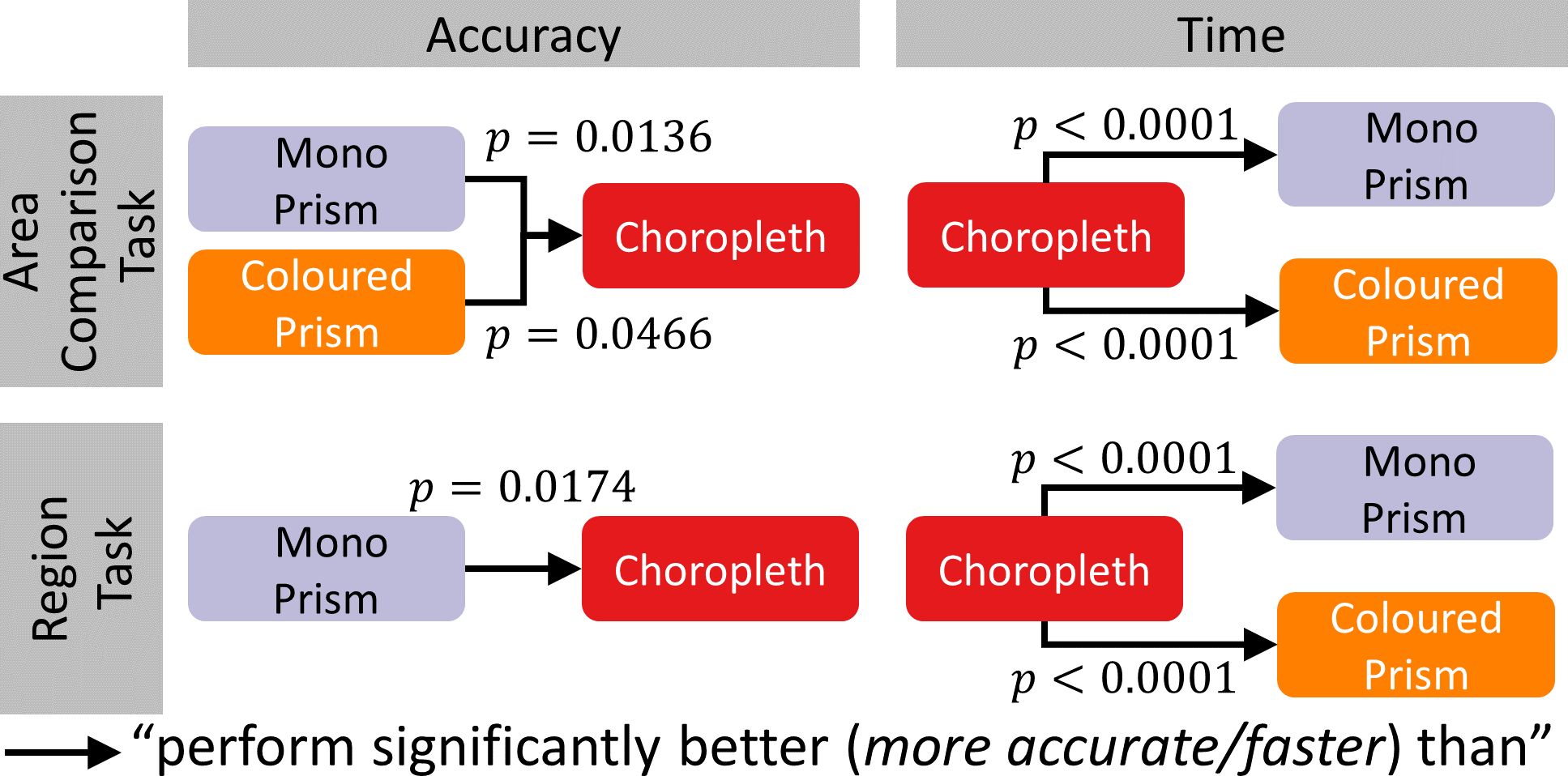}
	\caption{Study 1 --- Graphic depiction of statistic comparisons.}
	\label{fig:Exp-01-results-stats}
\end{figure}

\vspace{-0.5em}
\subsection{Results}
Since the distributions of dependent variables were skewed, we analyzed the \emph{square root} of both error rate (absolute difference, abbreviated to AbsDiff) and time.
\emph{Linear mixed modeling} was used to evaluate the effect of independent variables on the dependent variables~\cite{Bates2015}. All independent variables and their interactions were modeled as fixed effects. A within-subject design with random intercepts was used for all models. The significance of the inclusion of an independent variable or interaction terms were evaluated using the log-likelihood ratio. Tukey's HSD post-hoc tests were then performed for pair-wise comparisons using the least square means~\cite{Lenth2016}. Homoskedasticity and normality of the Pearson residuals were evaluated graphically using predicted vs residual and Q---Q plots respectively. 
Degree of freedom, $\chi^2$ and $p$ value for fixed effects were reported following~\cite[p.~601]{field2012discovering}. Accuracy (AbsDiff) and time with 95\% confidence of different visualisations are shown in Fig.~\ref{fig:Exp-01-results}.

\vspace{0.5em}
\noindent\textbf{Area-Comparison Task:} Three independent factors (visualisation or V, country or C and distance or D) and their interactions (V$\times$C, V$\times$D, C$\times$D and V$\times$C$\times$D) were modeled. 

The type of visualisation had a statistically significant effect on AbsDiff ($\chi^2(2)=9.2, p=.0101$). Post-hoc tests revealed that \textit{Monochrome Prism} and \textit{Coloured Prism} were statistically more accurate than \textit{Choropleth} (see Fig.~\ref{fig:Exp-01-results-stats}).

The type of visualisation also had a statistically significant effect on time ($\chi^2(2)=63.4, p<.0001$). Post-hoc tests revealed that \textit{Choropleth} was statistically faster than both \textit{Monochrome Prism} and \textit{Coloured Prism} (see Fig.~\ref{fig:Exp-01-results-stats}).

\vspace{0.5em}
\noindent\textbf{Region Task:} Two independent factors (V and C) and their interaction (V$\times$C) were modeled.

The type of visualisation had a statistically significant effect on AbsDiff ($\chi^2(2)=7.6, p=.0223$). Post-hoc tests revealed that \textit{Monochrome Prism} was statistically more accurate than \textit{Choropleth} (see Fig.~\ref{fig:Exp-01-results-stats}). The type of country had a marginally significant effect on AbsDiff ($\chi^2(1)=2.7, p=.0977$). Post-hoc tests revealed that participants tended to be more accurate in US than UK ($p=.0991$).

The type of visualisation also had a statistically significant effect on time ($\chi^2(2)=41.5, p<.0001$). Post-hoc tests revealed that \textit{Choropleth} was statistically faster than both \textit{Monochrome Prism} and \textit{Coloured Prism} (see Fig.~\ref{fig:Exp-01-results-stats}). The type of country had a marginally significant effect on time ($\chi^2(1)=3.4, p=.0661$). Post-hoc tests revealed that participants tended to spend more time on US than UK with $p=.0672$. However, the difference was small (see Fig.~\ref{fig:Exp-01-results}).  

Plotting the AbsDiff and time in the order of trials in the study revealed no obvious learning effect.

\begin{figure}
	\centering
	\includegraphics[width=0.9\columnwidth]{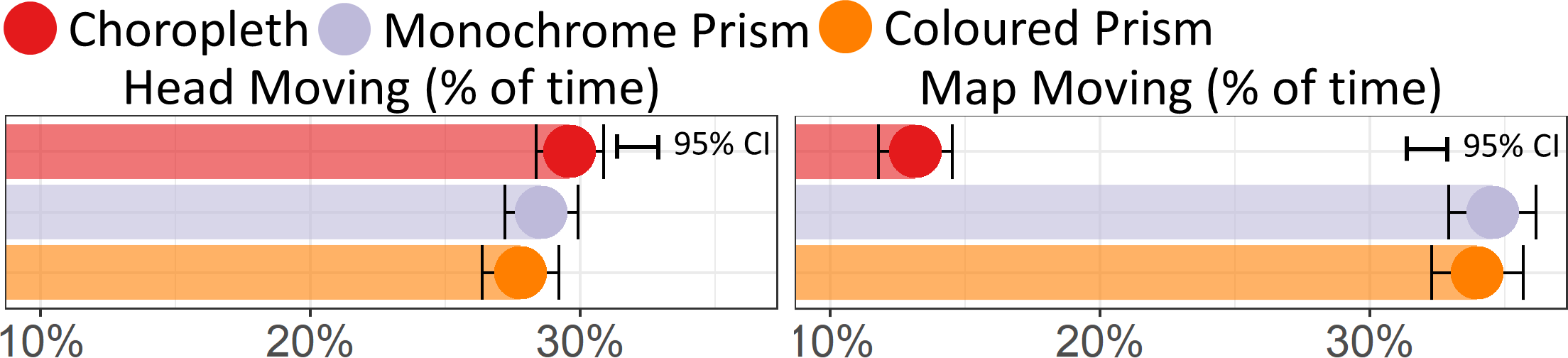}
	\caption{
	Time percentage in different movements.
	}
	\label{fig:Exp-01-interaction} 
\end{figure}

\vspace{0.5em}
\noindent\textbf{Interactions:} In order to investigate user interaction, we sampled every frame. If the head or the map moved more than 1~cm or rotated more than 5\textdegree, we classified it as an interaction, see Fig.~\ref{fig:Exp-01-interaction}. We accumulated the interaction time for each question and then normalised with respect to the total time spent on that question. The linear mixed model showed no statistically significant effect of visualisations on head movement, but on map movement ($\chi^2(2)=56.90, p<.0001$). Participants moved \emph{Monochrome Prism} and \emph{Coloured Prism} more often than \emph{Choropleth} (all statistically significant at $p<.0001$). 
\added{This is possibly because participants moved the maps more to deal with the occlusion in \emph{Monochrome Prism} and \emph{Coloured Prism}.}

\begin{figure}
    \vspace{0.8em}
	\centering
	\includegraphics[width=.8\columnwidth
	]{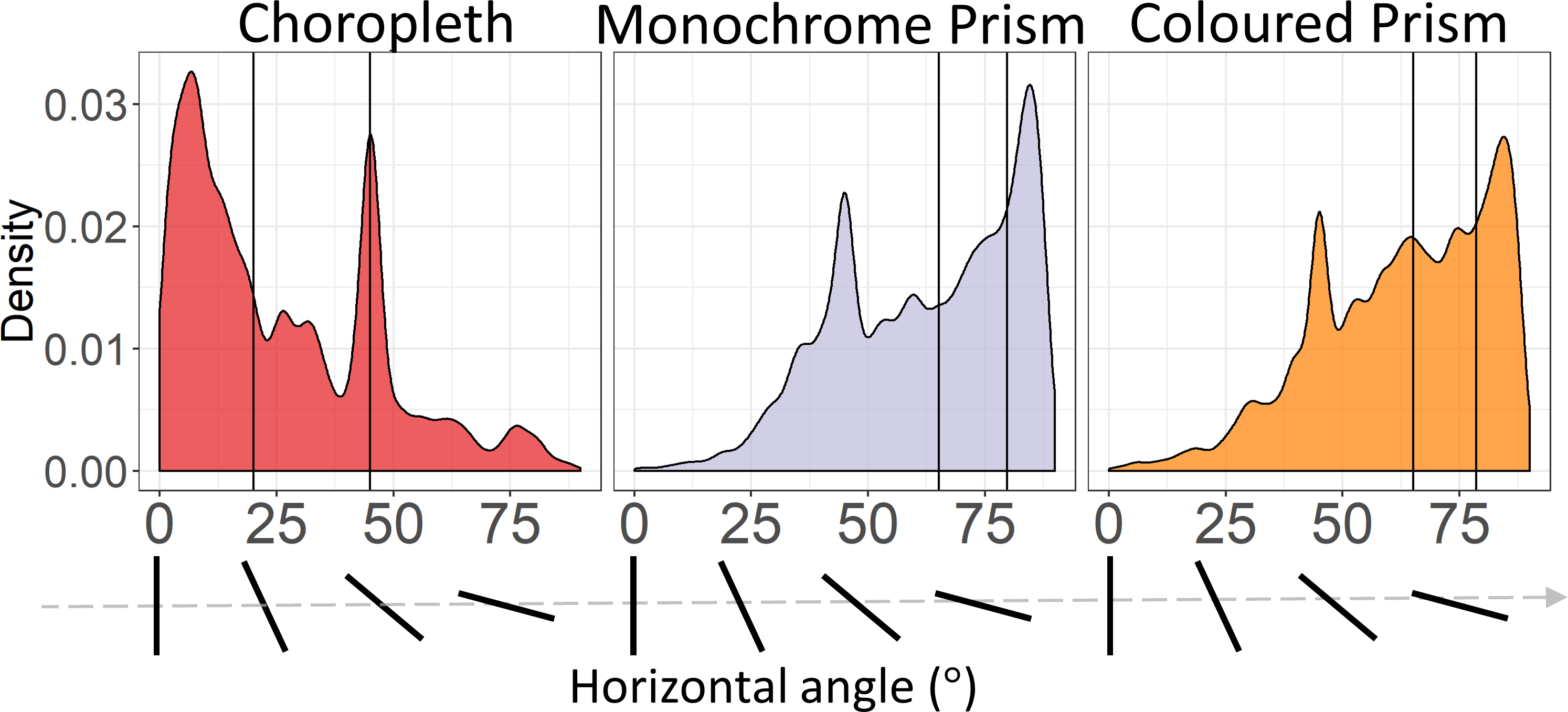}
	\caption{
	Tilt angle distribution for the different maps with median and third quartile lines.
	}
	\label{fig:Exp-01-angles}
\end{figure}

We also analyzed the tilt angle of the maps, i.e. the angle between the normal vector of the map plane and the horizontal plane (see Fig.~\ref{fig:Exp-01-angles}). 
Due to the non-normally distributed residuals, we used the Friedman test to compare the percentage of time spent with a view angle larger than 45\textdegree.
The test revealed a statistically significant effect ($\chi^2(2)=12.67, p=.0018$). As one might expect, participants spent statistically more time with a large tilt angle in \emph{Monochrome Prism} and \emph{Coloured Prism} than \emph{Choropleth}, all $p<.05$. In Fig.~\ref{fig:Exp-01-angles}, we see a peak of around 45\textdegree~for all visualisations, the probable reason is that maps were tilted to 45\textdegree~at the beginning of each question.

\begin{figure}
	\centering
	\includegraphics[width=\columnwidth]{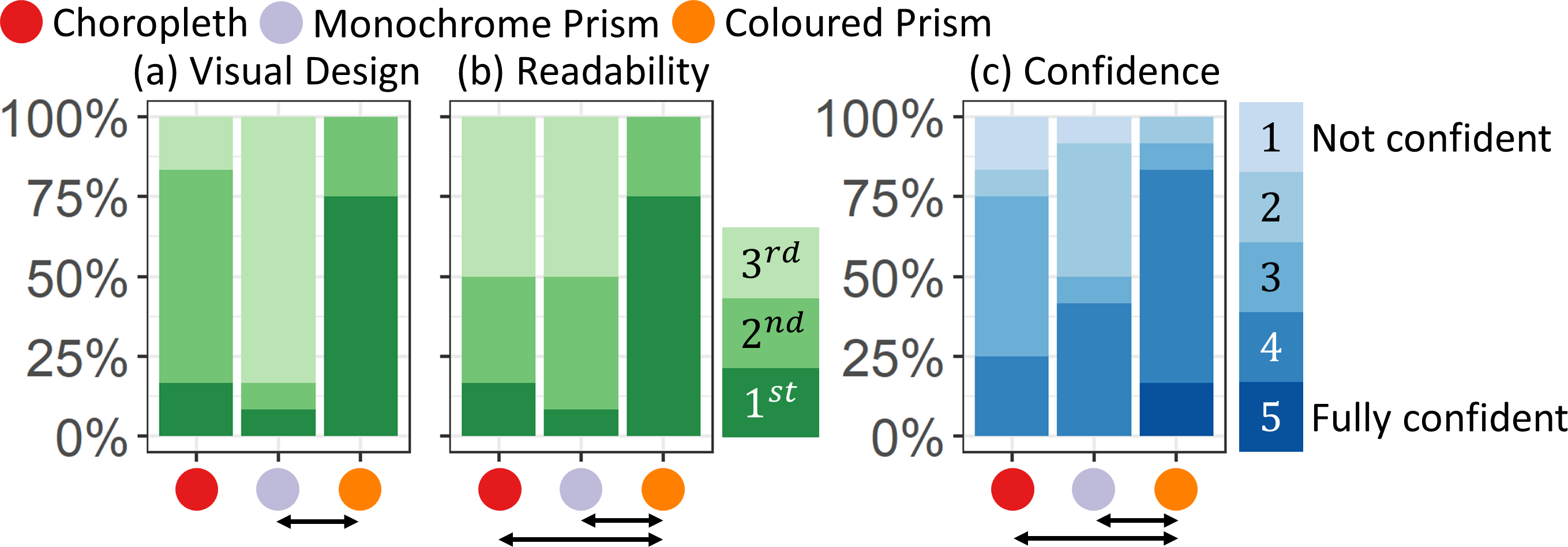}
	\caption{
	User preference:
	(a, b) ranking for each of the three visualisations; (c) confidence rating in five-point-Likert scale. Arrows indicate statistic comparisons with $p<.05$.
	}
	\label{fig:Exp-01-rankings}
\end{figure}

\vspace{0.5em}
\noindent\textbf{User preference:}
For \emph{visual design}, in Fig.~\ref{fig:Exp-01-rankings}(a), the strongest preference was for \emph{Coloured Prism}. 75\% participants ranked it as the best.
The Friedman test revealed a statistically significant effect of visualisation on preference ($\chi^2(2)=13.5, p=.0012$). The post-hoc tests found a statistically stronger preference for \emph{Coloured Prism} than \emph{Monochrome Prism} with $p=.0007$.  

For \emph{readability}, in Fig.~\ref{fig:Exp-01-rankings}(b), the strongest preference was again for \emph{Coloured Prism}. 75\% participants ranked it as the best. 
The Friedman test revealed a statistically significant effect of visualisation on preference ($\chi^2(2)=10.2, p=.0062$). The post-hoc tests found a statistically stronger preference for \emph{Coloured Prism} than \emph{Monochrome Prism} ($p=.0119$) and \emph{Choropleth} ($p=.0216$). 

For \emph{confidence}, in Fig.~\ref{fig:Exp-01-rankings}(c), the strongest preference was again for \emph{Coloured Prism} with 83.3\% rating it 4 or 5. The Friedman test revealed a statistically significant effect of visualisation on confidence ($\chi^2(2)=8.93, p=.0115$). The post-hoc tests found participants felt statistically more confident with \emph{Coloured Prism} than with \emph{Monochrome Prism} ($p=.0261$) and with \emph{Choropleth} ($p=.0261$). 

\vspace{0.5em}
\noindent\textbf{Strategies:}
Participant responses to the  questionnaire provided insight into their strategies for the two tasks.
They did not mention using special strategies for \emph{Choropleth}. For \emph{Monochrome Prism} and \emph{Coloured Prism}, participants tended to use similar strategies. For the area-comparison task, some  described how they computed the value of an occluded target. They searched for a geographic unit without occlusion adjacent to the target and read its value from the legend. Then they used this geographic unit as the ``yardstick'' to determine the target's value. For the region task,   two  strategies were described: \emph{Cut and fill}--imagine cutting the tall ones and using the volume to fill the lower ones; \emph{Divide and mix}--divide the set into two halves with similar size and  ``do the math''.

\vspace{0.5em}
\noindent\textbf{Key Findings:}
The results of this first comparison of 2D choropleth maps with 3D prism maps in an immersive environment accord with those of~\cite{Stewart:2010hi,Popelka:2018cy} using 2D monitors: participants are more accurate but slower with prism maps than choropleth maps.
This is in line with early  studies  showing that people are more able to discriminate differences in the size of marks than colour~\cite{cleveland1984graphical}. Specifically we found:
\begin{itemize}[leftmargin=3mm]
    \item \emph{Monochrome Prism} was more accurate than \emph{Choropleth}.
    \item \emph{Choropleth} was faster than both \emph{Monochrome} and \emph{Coloured Prism}.
    \item \emph{Coloured Prism} was more accurate than \emph{Choropleth} for the area-comparison task.
\end{itemize}
We also found that participants tended to move maps more with \emph{Monochrome} and \emph{Coloured Prism} than \emph{Choropleth} and tended to look from the side more with \emph{Monochrome} and \emph{Coloured Prism} than \emph{Choropleth}.
Comments such as \textit{``regions that were blocked were hard to read''} suggest that this may be because participants sought to find viewing angles to reduce occlusion and perspective distortion.

Participants felt more confident with \emph{Coloured Prism} than with \emph{Choropleth} and \emph{Monochrome Prism} and preferred \emph{Coloured Prism} to \emph{Choropleth} and \emph{Monochrome Prism} in terms of both visual design and readability.

\begin{figure}[t!]
    \centering
    \includegraphics[width=\columnwidth]{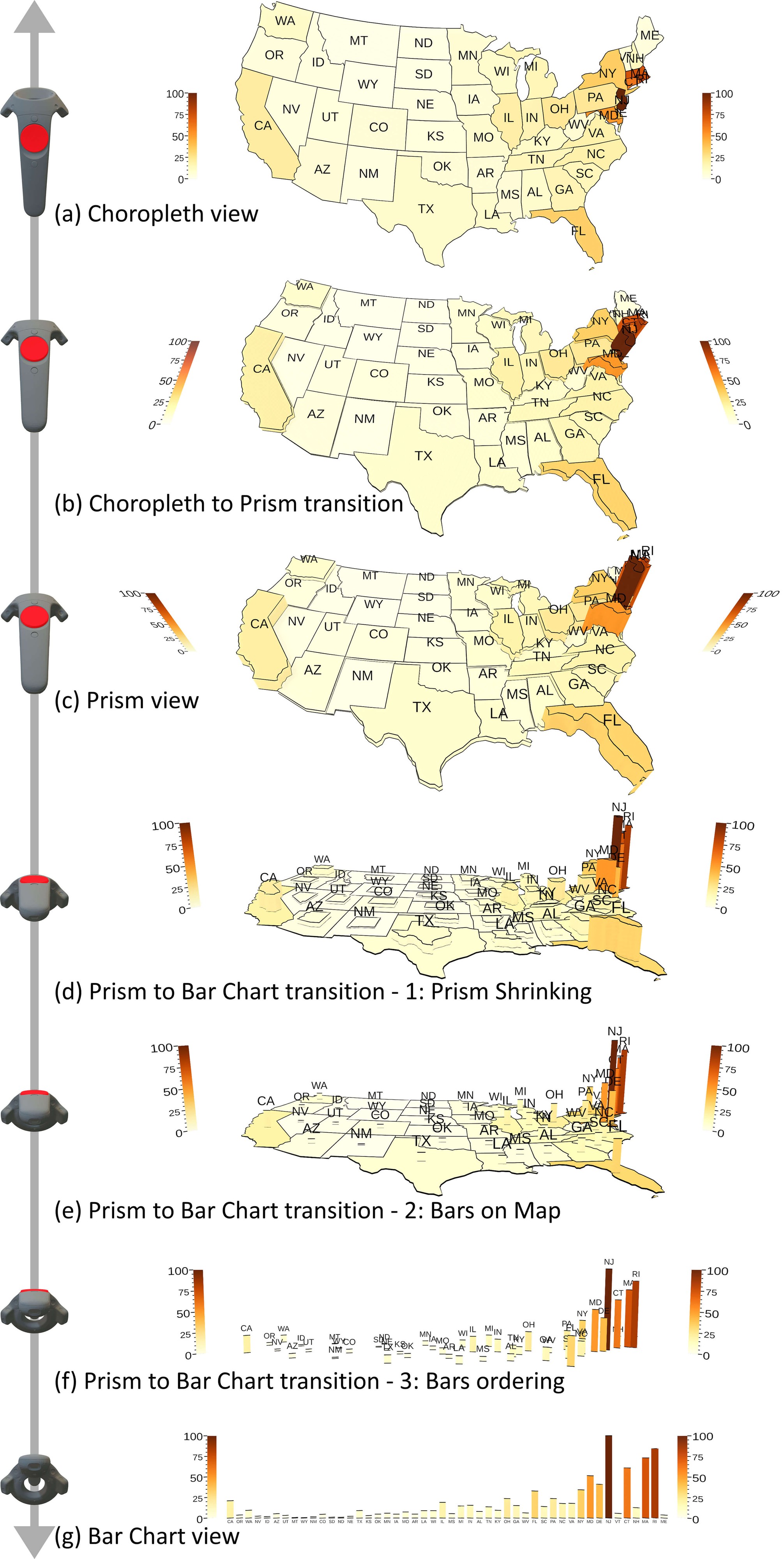}
    \caption{\emph{Tilt Map} combines (a) \emph{Choropleth}, (c) \emph{Prism}, and (g) \emph{Bar Chart} views with intermediate transitions. The current view is controlled by the tilt angle.
    }
    \label{fig:Exp-02-magic}
\end{figure}

\section{Tilt Map: A Hybrid 2D and 3D Method}
Study 1 found a trade-off between speed and accuracy for choropleth and prism maps in immersive environments.  
One could interpret this to suggest that choropleth provides a good ``overview" for quick comparisons between regions, however, an encoding that uses the third dimension to provide a spatial mapping of value (such as prism) is better for precise reading. We therefore felt that a hybrid combination of the two views might have advantages over either on its own. Furthermore, while using height to encode data in prism maps improved accuracy, their 3D nature led to issues with occlusion and perspective distortion. We therefore thought that it would be beneficial to provide a complementary 2D bar chart view.

Complementary visualisations are commonly combined in two ways. The Tableau data visualisation software, for example, allows for the creation of tabbed displays where users can switch between different views that take up the whole screen.  Such tools also support the creation of dashboard displays with tiled visualisations positioned side-by-side.  Usually in such a tiled display each individual view shows a different subset or different dimensions of the data.  It is also possible to use \emph{perceptually complementary} displays~\cite{chang2018evaluation}. These show the same data using different visualisation techniques, each of which is better suited to specific analysis tasks.

A third option for combining views, however, is to unify the best aspects of each view into a single conceptual display. \added{Indeed, the \emph{Coloured Prism} can be considered as a hybrid representation.}
Viewed directly from above, it appears like a choropleth map, but not perfectly.  Linear perspective means that the sides of prisms at the periphery of the view are visible and that areas that are closest to the viewer will appear larger, exaggerating their quantity.  Furthermore, lighting effects can cause shadows to be cast across lower areas.  From the side (which  Fig.~\ref{fig:Exp-01-angles} shows was a common viewing angle in Study~1) the \emph{Prism} map appears like a bar chart.  But it is a 3D bar chart that suffers from occlusion and perspective distortion.

Using the ``magic'' of virtual reality, however, we can correct for all of these issues.  As illustrated in Fig.~\ref{fig:Exp-02-magic}, we can dynamically morph the 3D model in response to tilt angle.  When viewed from directly above we can completely flatten the model so it is precisely a 2D choropleth map (Fig. \ref{fig:Exp-02-magic}a).  As the user rotates the model (a natural gesture using the 6DOF tracked VR controller), we can scale the height of the prisms directly with view angle (Fig. \ref{fig:Exp-02-magic}b), until the visualisation is a true prism map at about $45^\circ$ (Fig. \ref{fig:Exp-02-magic}c).  As the user continues to rotate the model we can further adapt the display, flattening the prisms to eliminate foreshortening and sliding them sideways to remove occlusion (Fig. \ref{fig:Exp-02-magic}d-f).  Viewed from $90^\circ$ the view that we show is precisely a 2D bar chart (Fig. \ref{fig:Exp-02-magic}g).  
It is worth noting that a couple of the stages are reminiscent of other thematic  map visualisation techniques.  In particular, Fig. \ref{fig:Exp-02-magic}d resembles a non-contiguous cartogram 
and Fig. \ref{fig:Exp-02-magic}e resembles bars on a map (recently popularised by the 3D map feature in MS Office 365). 

To differentiate this morphing hybrid from the other conditions of our next study, we refer to it as a ``\emph{Tilt Map}''.

\subsection{Tilt Map Design Considerations}
\noindent\textbf{Colouring:} A difficult design decision was whether to colour the \emph{Prism} view. While our previous study suggested possible benefits to using a monochrome prism map, we decided that the visual continuity provided by preserving the colouring of the choropleth view outweighed the potential downside. 

\vspace{0.5em}
\noindent\textbf{User Control:}
In all conditions, users were able to freely move and rotate maps by moving the controller inside the map boundary and ``grasping'' the controller's trigger.  From the time it is grasped until the time it is released the map's position and angle relative to the ground is ``latched'' to the controller.
In the \emph{Tilt Map} we chose to couple the angle of the map relative to the ground, in the vertical plane of the user's view vector, to the type of visualisation shown. 

\vspace{0.5em}
\noindent\textbf{Choice of transition angle intervals:}
Transitions from \emph{Choropleth} view, to \emph{Prism} view, to \emph{Bar Chart} view, occur at key angles within a $90^\circ$ arc of angle relative to the ground.
The precise angles triggering the transitions, indicated in the figure to the right, were informed by our observations of angles at which users tended to view the different map techniques in Study 1, see Fig.~\ref{fig:Exp-01-angles}.
\added{Fig.~\ref{fig:teaser} (right) shows labels (a--g) that correspond to the viewing angles in Fig.~\ref{fig:Exp-02-magic}.}
Transitions are not sudden but continuously controlled by tilt angle so that the user sees a smooth animated transitions as they tilt the map.

\vspace{0.5em}
\noindent\textbf{Projection order of prisms to bars:}
A key design decision for the transition from \emph{Prism} view to \emph{Bar Chart} was to order the bars corresponding to the left-to-right order of the area centres in the view plane of the \emph{Prism} view.  In other words, a straightforward projection of the regions into the left-to-right view axis.  Thus, the order of the bars depends upon the horizontal view angle. If the user was viewing the \emph{Prism} view with the southern edge of the map closest to them at the time of transition to the bar chart, the bars would be ordered west to east; if the eastern edge of the map was closest to them, the bars would be ordered south to north; and so on.
\added{This design choice minimised the relative movement of the bars when transitioning, which we felt was a relevant feature and was likely to minimize occlusion. Other orderings, e.g. from smallest value to largest, are also possible, but would lead to larger movements during transition.}

\vspace{0.5em}
\noindent\textbf{Legend/axis positions and orientation:}
Following immersive visualisation best practice, text labels in all views are billboarded.  That is, they reorient dynamically as the view angle changes such that they always face the user directly.

For the \emph{Choropleth} view it is essential to show a legend indicating the mapping of data value to area colour.  For the \emph{Prism} and \emph{Bar Chart} views it is also necessary to show axes indicating mapping from data value to prism/bar height.  For \emph{Tilt Map} we combine each colour legend and axis into a single annotation.  It is important that this annotation be clearly visible regardless of the view angle, however, the axes in the \emph{Prism} view cannot be bill-boarded, since they must always lie in the height axis.  In the \emph{Choropleth} view it is natural that the legend lie in the plane of the map.  Thus, another transition occurs for these legend/axis annotations: the axis rotates from the map plane to the height axis as the \emph{Tilt Map} transitions from the \emph{Choropleth} to the \emph{Prism} view.  
In \emph{Choropleth} and \emph{Prism} views there are four of these legend/axis annotations, positioned at the four sides of the map.  In the final transition to the \emph{Bar Chart} view, the top and bottom axes are removed leaving only axes on the left and right of the bar chart.

\vspace{-0.5em}
\section{Study 2: Tilt Map Evaluation}

We conducted a second user study in order to evaluate \emph{Tilt Map}. We wished to compare it to more traditional ways to combine a choropleth, prism and bar chart view.
We compared \emph{Tilt Map} to:

\vspace{0.5em}
\noindent\textbf{Side-By-Side:}
We placed three views following the transition order of \emph{Tilt Map}, i.e. from left to right: \emph{Choropleth} view (left), \emph{Prism} view (centre) and \emph{Bar Chart} view (right). All three views were at the same distance to the viewer. The \emph{Prism} view was immediately in front of the viewer, while the \emph{Choropleth} and \emph{Bar Chart} views were positioned in an egocentric layout, 80\textdegree~anticlockwise and clockwise respectively.
Participants were free to rearrange all three views independently, but the layout was reset at the beginning of each trial.

\vspace{0.5em}
\noindent\textbf{Toggle:}
Participants could switch the view by tapping the left or right half of the touchpad on the controller to cycle forward or backward. The initial view was \emph{Choropleth} view, and consistent with \emph{Tilt Map}, the forward-toggle order was \emph{Choropleth} -- \emph{Prism} -- \emph{Bar Chart}. 

\vspace{-0.5em}
\subsection{Experiment}
\added{Following the same structure of Sec.~\ref{sec:exp-1}, in this subsection, we first introduce the \emph{tasks} of the user study and the way we create task \emph{data}. We then report details of the user study including: \emph{experimental set-up}, \emph{design and procedure}, \emph{participants} and \emph{measures}.}

\vspace{0.5em}
\noindent\textbf{Tasks:} 
We evaluated \emph{Tilt Map}, \emph{Side-By-Side} and \emph{Toggle} using the \emph{Region Task} from the first user study. 
We kept the coefficient of variation (CV) within 40\%$-$60\% for all tasks.
The answers to the generated tasks evenly fell into three different ranges: 20$-$40, 40$-$60 and 60$-$80. 
We did not use the area-comparison task as the \emph{Bar Chart} view made this task too easy. We used 5 contiguous states as the target region for the US and 10 contiguous areas for the EU.

\vspace{0.5em}
\noindent\textbf{Data:} 
We again tested using two datasets. We used the same US data as in the first user study. We chose not to use the  UK data because its large size made the bar chart view too wide.
Instead we generated data based on Europe (EU) data \cite{europedata}. We used the first level of Nomenclature of Territorial Units for Statistics (NUTS). After removing a few areas geographically distant from Europe, the dataset contained 116 geographic areas. 
Again, the original population density data was highly skewed
\added{with only very few areas with high population densities. To  ensure the
map visualisations used the entire range of colour or height variation
(see Sec.~\ref{sec:exp-1}~---~Data)}, here we applied a fourth root transformation. 
We again generated different data for each question and kept Moran's \textit{I} the same as the original data.

\vspace{0.5em}
\noindent\textbf{Experimental Set-up:}
PC and headset were as per the first study. All maps were created on top of a transparent quadrangle of 1$\times$1m. For \emph{Tilt Map} and \emph{Toggle}, the initial views were \emph{Choropleth} view with 0\textdegree~tilting. For \emph{Side-By-Side}, the \emph{Choropleth} and \emph{Bar Chart} views were placed perpendicularly to the horizon plane and the \emph{Prism} view was tilted to 75\textdegree. The initial views of \emph{Tilt Map} and \emph{Toggle} were placed the same as in the first study, i.e. 0.6~m in front of the participants' eye position and 0.1~m below it. For \emph{Side-By-Side}, to avoid overlap among views, the distance was enlarged to 0.9~m.

\vspace{0.5em}
\noindent\textbf{Design and Procedure:} 
The experiment was within-subjects: 18 participants $\times$ 3 visualisations $\times$ 1 task $\times$ 2 datasets $\times$ 3 answer ranges $\times$ 2 repetitions = 648 responses (36 responses per participant) and lasted one hour and ten minutes on average. Latin square design was used to balance the order of visualisations.

The study procedure was similar to the first study but with two modifications:

\noindent\emph{(1) A pre-study training was added.} Based on the participants recruited in the first study, we expected most participants to have limited experience with VR. 
Therefore, we recruited a participant without VR experience to do a pilot study. The pilot study revealed some initial difficulty with VR interactions. Thus, we designed a pre-study training which instructed participants in basic VR interactions, including repositioning, rotating and tilting a 2D flat world map. Participants were given unlimited time to become familiar with these interactions. The pre-study training usually took 5--10 minutes. 

\noindent\emph{(2) Questions were added to the post-hoc questionnaire.} We asked participants to rate the usefulness of each view within the combined visualisations with a five-point Likert scale, i.e. \emph{Choropleth} view, \emph{Prism} view and \emph{Bar Chart} view. We also asked participants to rate the usefulness of the continuous transition in \emph{Tilt Map} with a five-point Likert scale and explain their reasons for the rating.

\vspace{0.5em}
\noindent\textbf{Participants:}
We recruited 18 new participants (5 female, 13 male) from our university. All had normal or corrected-to-normal vision and included students and researchers. 
12 participants were aged between 20$-$30, 5 between 30$-$40, and 1 over 40. VR experience varied: 10 participants had less than 5~h of prior VR experience, 3 had 6$-$20~h, and 5 more than 20~h. 

\vspace{0.5em}
\noindent\textbf{Measures:}
In addition to real-time recording of participant's head, controller, map position and map rotation, we recorded the view the participant was looking at.
\begin{figure}
	\centering
	\includegraphics[width=1\columnwidth]{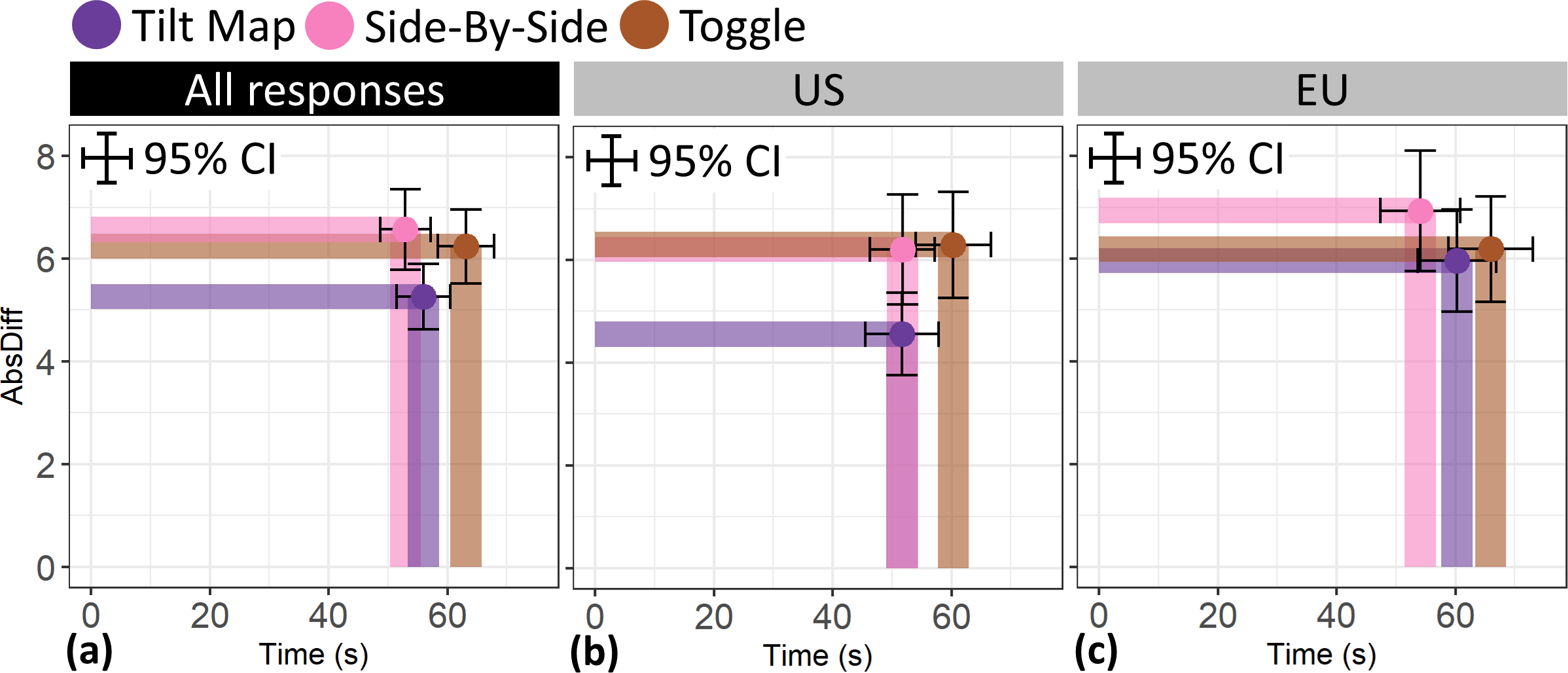}
	\caption{Study 2 --- Accuracy and response time with 95\% confidence intervals (\textit{AbsDiff} = absolute difference between participants' answers and the correct answers.)}
	\label{fig:Exp-02-results}
\end{figure} 
\begin{figure}
    \vspace{1em}
    \centering 
	\includegraphics[width=0.95\columnwidth]{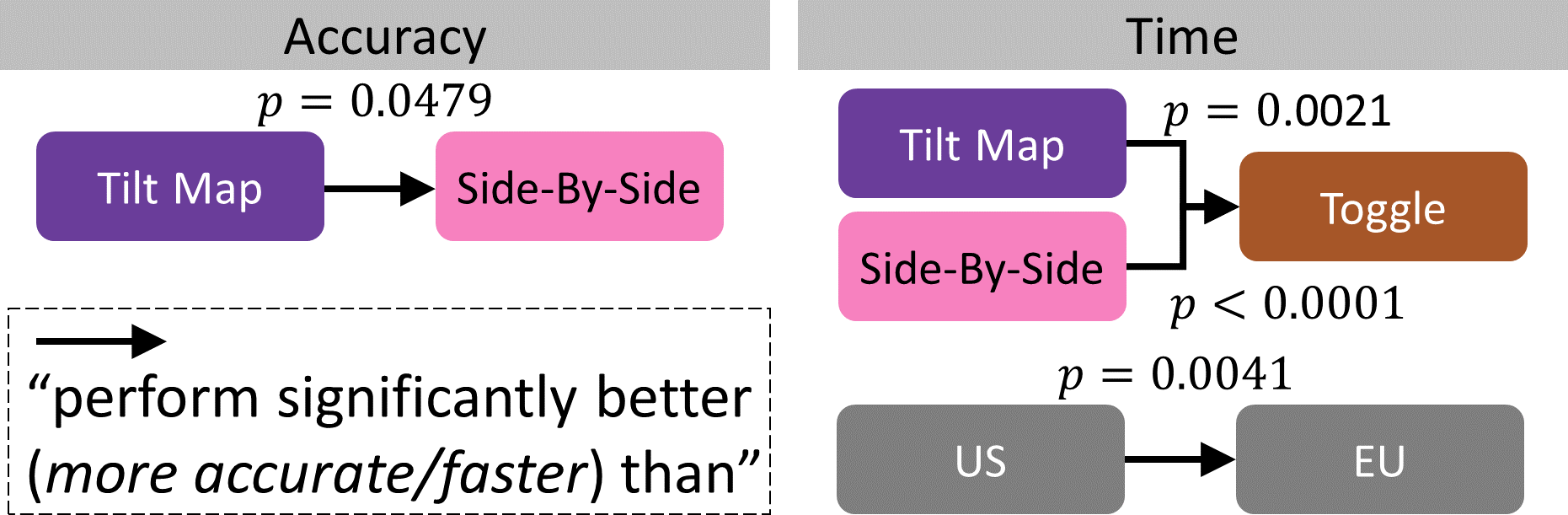}
	\caption{Study 2 --- Graphic depiction of statistic comparisons with $p<.05$.}
	\label{fig:Exp-02-results-stats}
\end{figure}

\subsection{Results}
As in the first study, we used \emph{linear mixed modeling} to check for significance and applied Tukey's HSD post-hoc tests to conduct pairwise comparisons for the \emph{square root} of AbsDiff and time. Two independent factors (visualisation and country) and their interaction (visualisation$\times$country) were modeled. Accuracy (AbsDiff) and time with 95\% confidence of different visualisations were presented in Fig.~\ref{fig:Exp-02-results}.

The type of visualisation had a statistically significant effect on AbsDiff ($\chi^2(2)=6.5, p=.0393$). Post-hoc tests revealed that \emph{Tilt Map} was statistically more accurate than \textit{Side-By-Side} (see Fig.~\ref{fig:Exp-02-results-stats}). The type of country also had a marginally significant effect on AbsDiff ($\chi^2(1)=3.6, p=.0569$). Post-hoc tests revealed that participants were marginally more accurate in US tasks than EU tasks.

The type of visualisation had a statistically significant effect on time ($\chi^2(2)=23.1, p<.0001$). Post-hoc tests revealed that \textit{Toggle} was statistically slower than both \textit{Tilt Map} and \textit{Side-By-Side} (see Fig.~\ref{fig:Exp-02-results-stats}). The type of country also had a statistically significant effect on time ($\chi^2(2)=8.3, p=.0040$). Post-hoc tests revealed that participants were statistically faster in US tasks than EU tasks (see Fig.~\ref{fig:Exp-02-results-stats}).

Since one task and one dataset were identical in the two studies, we were able to compare the effectiveness of the hybrid visualisations in Study~2 with the single visualisations used in Study~1 for the US region task. \added{Following mixed designs modeling~\cite[Chapter~14]{field2012discovering},} we used linear mixed model to check statistical significance and applied Tukey's HSD post-hoc tests to conduct pairwise comparisons. 
We found a statistically significant effect of visualisations on AbsDiff ($\chi^2(5)=18.3, p=.0026$). In addition to the findings in Fig~\ref{fig:Exp-02-results-stats}, we found that, \emph{Tilt Map} was statistically more accurate than \emph{Choropleth} and \emph{Coloured Prism}, with $p=.0024$ and $p=.0082$ respectively. 
We also found a statistically significant effect of visualisations on time ($\chi^2(5)=53.2, p<.0001$). \emph{Choropleth} was faster than all other visualisations and \emph{Toggle} was slower than all other visualisations, all $p<.05$. We found that \emph{Tilt Map} had very similar accuracy and time to \emph{Monochrome Prism}. 
This suggests that the choice to colour the prism view in \emph{Tilt Map} did not degrade performance.

Again, we plotted the AbsDiff and time against task order: again we found no obvious learning effect.

\begin{figure}
	\centering
	\includegraphics[width=1.0\columnwidth]{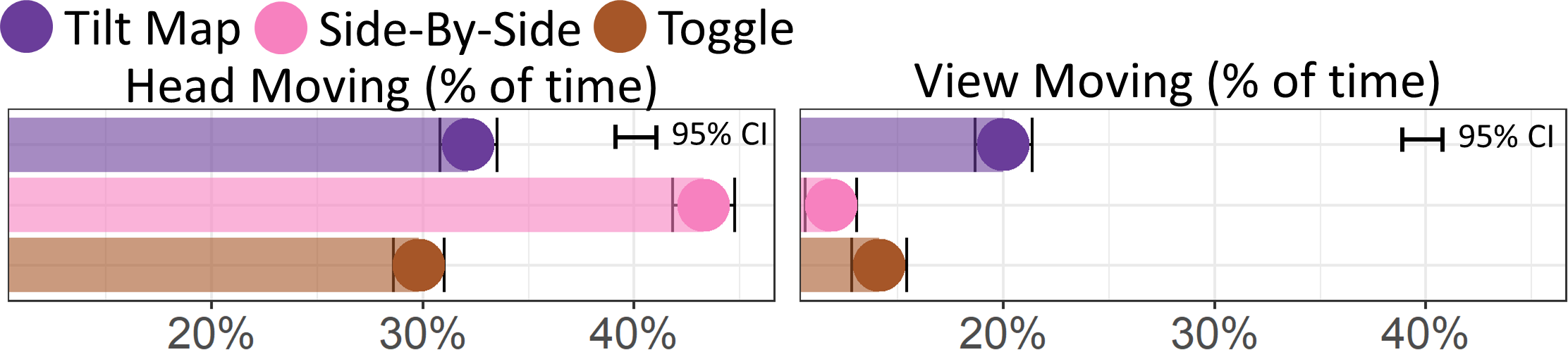}
	\caption{Time percentage in different movements.}
	\label{fig:Exp-02-interactions}
\end{figure}

\begin{figure}
    \vspace{1em}
	\centering
	\includegraphics[width=1.0\columnwidth]{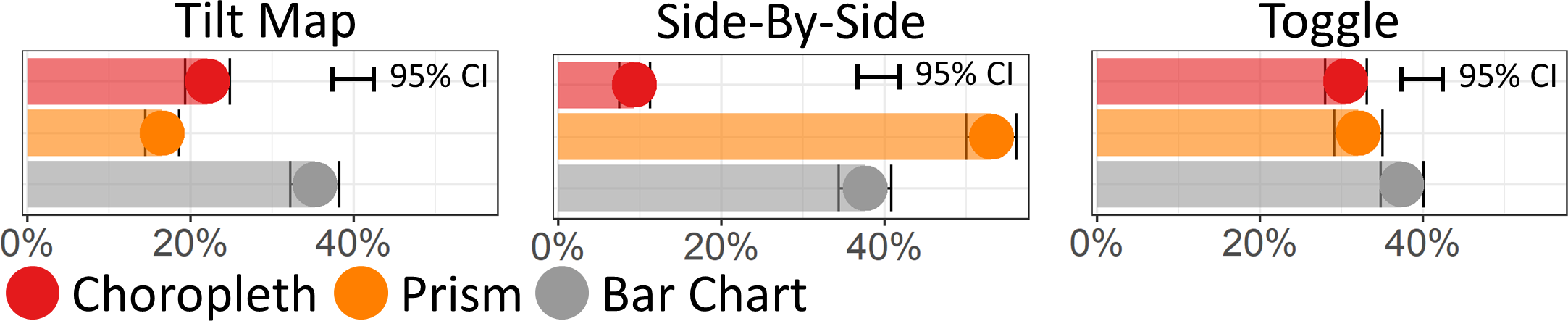}
	\caption{Time percentage in each main view with 95\% confidence interval.}
	\label{fig:Exp-02-view-percentage}
\end{figure}

\vspace{0.5em}
\noindent\textbf{Interactions:} 
Linear mixed model was used to analyse the differences in the way participants interacted with the three visualisations. Unsurprisingly, participants spent a  statistically greater percentage of time moving their heads in \emph{Side-By-Side} than \emph{Tilt Map} and \emph{Toggle}. Participants also spent statistically more time manipulating the views in \emph{Tilt Map} than \emph{Side-By-Side} and \emph{Toggle}. Within each visualisation, participants spent statistically more time moving their head than moving the visualisation (all $p<.001$).

We investigated the time spent in each main view (i.e. \emph{Choropleth} view, \emph{Prism} view and \emph{Bar Chart} view). In Fig.~\ref{fig:Exp-02-view-percentage}, we can identify different patterns for different visualisations. For \emph{Tilt Map}, participants spent significantly more time in \emph{Bar Chart} view than \emph{Choropleth} view or \emph{Prism} view. For \emph{Side-By-Side}, participants spent significantly more time in \emph{Bar Chart} view and \emph{Prism} view than \emph{Choropleth} view, while time spent in \emph{Prism} view was more than \emph{Bar Chart} view. For \emph{Toggle}, participants spent time almost evenly in each of the three views.

In the \emph{Tilt Map}, participants could continuously morph between the three main views. Participants spent on average 13.5\% of time in the intermediate state between \emph{Choropleth} and \emph{Prism} (see Fig.~\ref{fig:Exp-02-magic}(b)), they also spent similar time (avg. 12.7\%) in the intermediate state between \emph{Prism} and \emph{Bar Chart} (see Fig.~\ref{fig:Exp-02-magic}(d,e,f) and discussed further in \textbf{Strategies}).

To tilt the \emph{Tilt Map} most participants grasped it holding the controller vertically as per Figures \ref{fig:teaser} and \ref{fig:Exp-02-magic}. 
\noindent\begin{minipage}{0.65\columnwidth}
However, three of the 18 participants grasped it holding the controller horizontally (like a motorcycle throttle), as per the figure to the right.  This was unexpected and interesting as it allows for a greater range of rotation.  There were not enough participants using this strategy to determine with any statistical significance if it improved performance, but it made us reflect that a handle affordance could be added to the \emph{Tilt Map} to encourage such a grasp.\unskip\parfillskip 0pt \par
\end{minipage}
\begin{minipage}{0.3\columnwidth}
\centering
\vspace{1mm}
\includegraphics[width=\columnwidth]{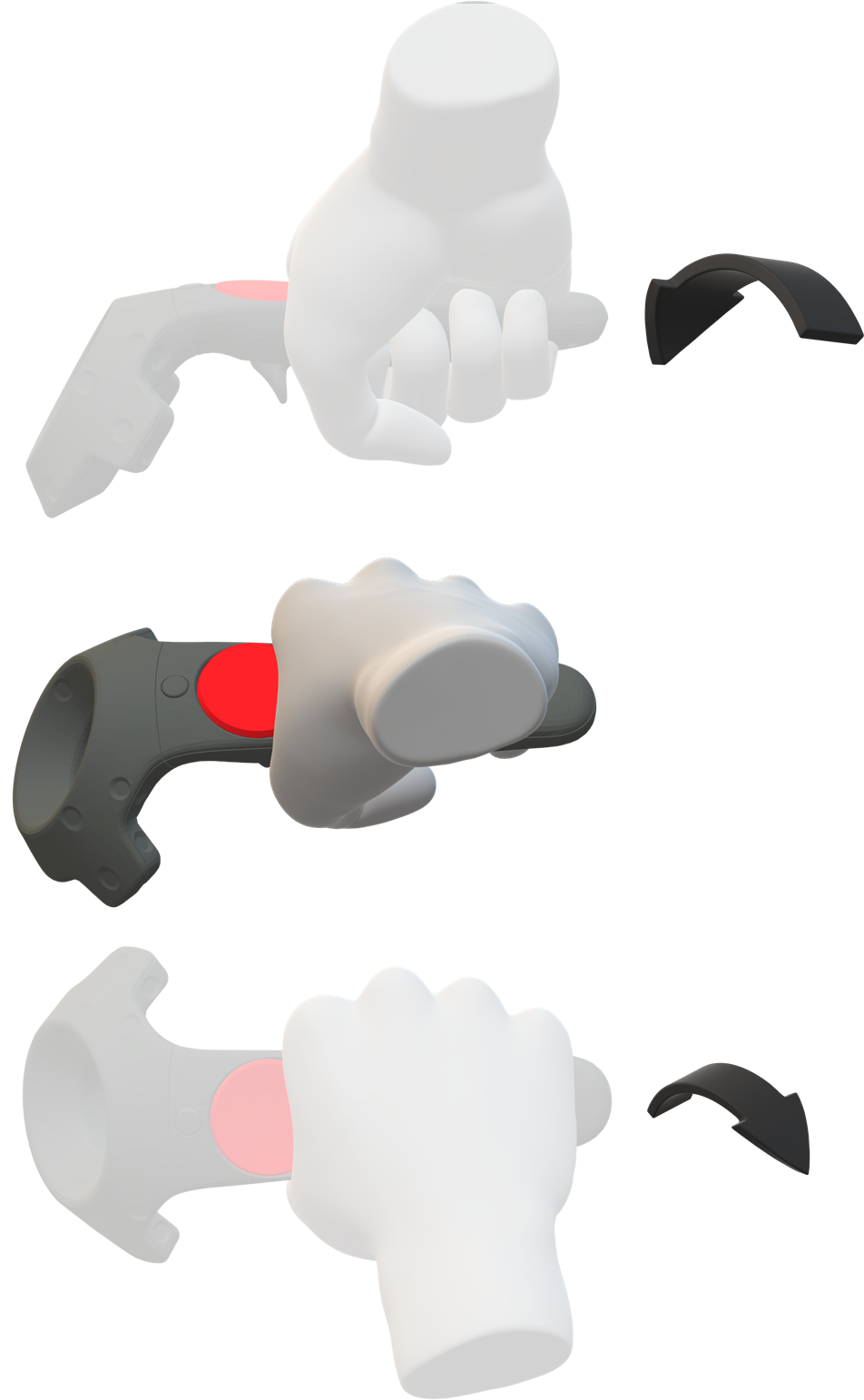} 
\vspace{-2mm}
\end{minipage}

\begin{figure}
	\centering
	\includegraphics[width=0.95\columnwidth]{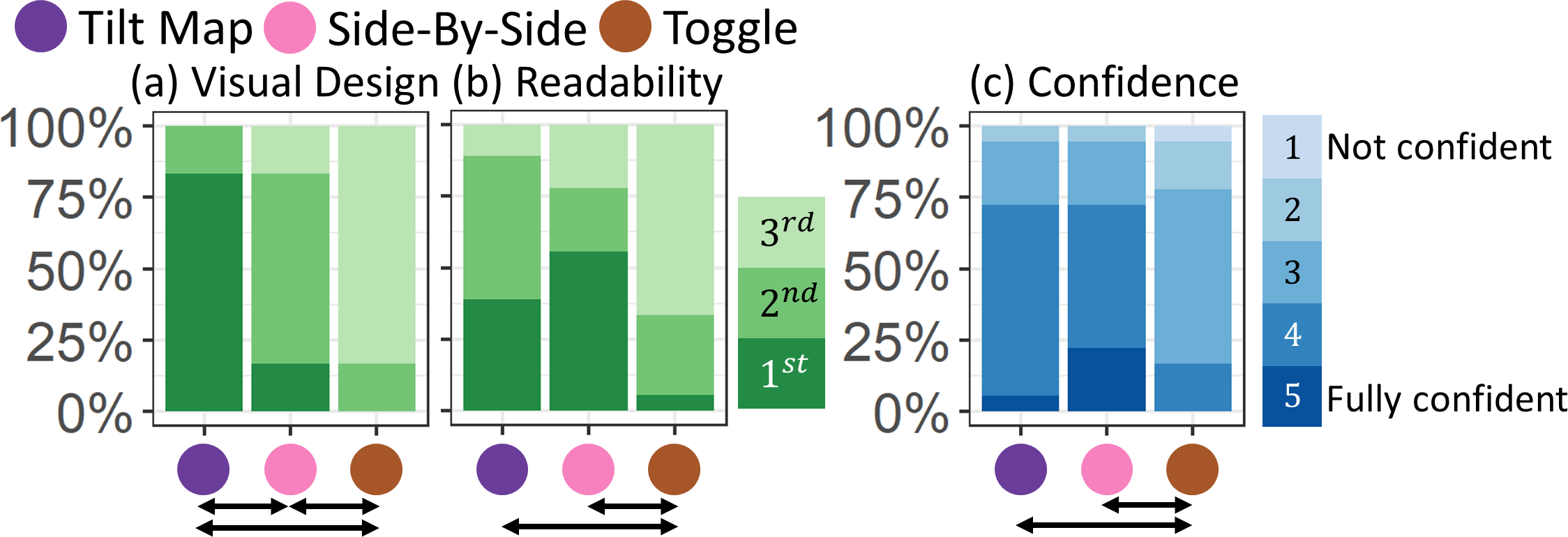}
	\caption{
	User preference:
	(a, b) ranking for each of the three visualisations; (c) confidence rating in five-point-Likert scale. Arrows indicate statistic comparisons with $p<.05$.
	}
	\label{fig:Exp-02-rankings}
\end{figure}

\vspace{0.5em}
\noindent\textbf{User Preference:}
For \emph{visual design}, in Fig.~\ref{fig:Exp-02-rankings}(a), the strongest preference was for \emph{Tilt Map} with 83.3\% participants ranking it as the best. 
The Friedman test revealed a significant effect of visualisation on preference ($\chi^2(2)=25, p<.0001$). The post-hoc tests confirmed \emph{Tilt Map} was statistically preferred to \emph{Side-By-Side} ($p=.0333$) and \emph{Toggle} ($p<.0001$). \emph{Side-By-Side} was statistically preferred to \emph{Toggle} ($p=.0333$).
 
For \emph{readability}, in Fig.~\ref{fig:Exp-02-rankings}(b), \emph{Toggle} was least preferred with only 5.6\% voting it as the best.
The Friedman test revealed a significant effect of visualisation on preference ($\chi^2(2)=10.11, p<.0001$). The post-hoc tests confirmed \emph{Tilt Map} and \emph{Side-By-Side} were statistically more preferred than \emph{Toggle} with $p=.0209$ and $p=.0128$, while there is no statistical difference between \emph{Tilt Map} and \emph{Side-By-Side}.

For \emph{confidence}, in Fig.~\ref{fig:Exp-02-rankings}(c), participants felt confident with \emph{Tilt Map} and \emph{Side-By-Side} with both 72.2\% rating them 4 or 5. 
The Friedman test revealed a statistically significant effect of visualisation on confidence ($\chi^2(2)=13.3, p=.0013$). The post-hoc tests found participants felt statistically more confident with \emph{Tilt Map} and with \emph{Side-By-Side} than with \emph{Toggle} with $p=.0076$ and $p=.0029$ respectively.

Additionally, we asked participants to rate the importance of each individual view (i.e. \emph{Choropleth}, \emph{Prism} and \emph{Bar Chart} views) with a five-point Likert scale. Participants rated the three views as equally important: 61.1\% rated \emph{Choropleth} view 4 or 5 (i.e. obviously useful), 72.2\% rated \emph{Prism} view obviously useful and 61.1\% rated \emph{Bar Chart} view obviously useful. The small difference was not statistically significant. 

We also asked participants to rate the usefulness of the morphing (seeing intermediate states of view changing) in the \emph{Tilt Map} with a five-point-Likert scale. 77.8\% rated the usefulness of morphing obviously useful. We also asked participants to explain their rating. Most participants commented that the tilting interaction was ``intuitive'' in VR. Some detailed comments revealed:
\begin{itemize}[leftmargin=*]
	\item The \emph{Prism Shrinking} transition reduced occlusion and the final stage, \emph{Bars on Map} transition, was found to be very useful (5 participants mentioned this). 
	\item The transition from the \emph{Prism} view to the \emph{Bar Chart} view helped them locate the position of areas in the \emph{Bar Chart} view (8 participants mentioned this).
\end{itemize}

\begin{figure}[t!]
    \centering
    \includegraphics[width=\columnwidth]{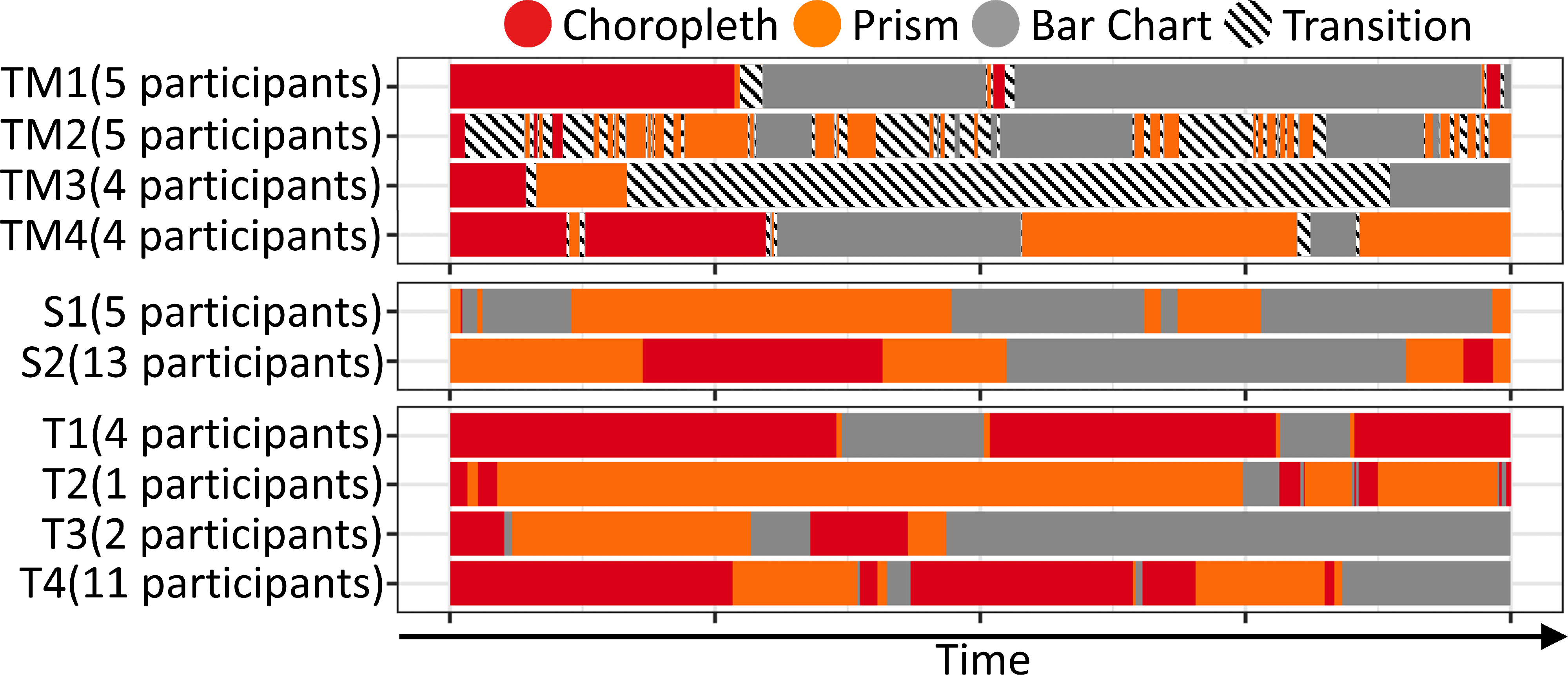}
    \caption{Demonstration of representative strategies for different visualisations. (TM1, TM2, TM3, TM4) for Tilt Map, (S1, S2) for Side-By-Side, (T1, T2, T3, T4) for Toggle.}
    \label{fig:Exp-02-strategies}
\end{figure}

\vspace{0.5em}
\noindent\textbf{Strategies:}
To better understand how participants interacted with the visualisations, we performed an exploratory analysis of the sequence of views used by participants  in order to provide an initial identification of common strategies used for each visualisation (see Fig.~\ref{fig:Exp-02-strategies}).

\noindent For \emph{Tilt Map,} \\
\noindent\emph{TM1} spent most time with \emph{Choropleth} and \emph{Bar Chart} views. \\
\noindent\emph{TM2} switched between views frequently. \\
\noindent\emph{TM3} spent a significant time on transitions. \\
\noindent\emph{TM4} spent similar time in each view.

\noindent For \emph{Side-By-Side,}\\
\noindent\emph{S1} spent most time with the \emph{Prism} and \emph{Bar Chart} views.\\
\noindent\emph{S2} spent similar time in each view.

\noindent For \emph{Toggle,}\\
\noindent\emph{T1} spent most time with the \emph{Choropleth} and \emph{Bar Chart} views.\\
\noindent\emph{T2} spent most time with the \emph{Prism} view.\\
\noindent\emph{T3} spent most time with the \emph{Prism} and \emph{Bar Chart} views.\\
\noindent\emph{T4} spent similar time in each view.

From our observations and the questionnaire we further identified a specific strategy in S1: 3 participants repositioned \emph{Prism} and \emph{Bar Chart} views to allow both of them in their field of view (vertically aligned). For the other participants who used S1, instead of changing the layout, they rotated their heads to switch the view.

\added{We found that the strategies used by participants were relatively consistent in \emph{Side-By-Side} and \emph{Toggle}, while strategies were much more diverse in \emph{Tilt Map}.} 

\vspace{0.5em}
\textbf{Feedback:}
The final section of the study allowed participants to give feedback on the pros and cons of each design. Qualitative analysis of these comments revealed (overall):

\emph{Tilt Map} was found to be \textit{``intuitive and cool''}. Most participants \textit{``enjoyed using it as it `gratified' in a way''} and \textit{``it is easy to control which view I want''}. Some found the transition and the intermediate states helpful. However, some also mentioned they felt it \textit{``a bit tiring''} and the transition from the \textit{Bars on Map} to \textit{Bar Chart} to be \textit{``a bit too fast''}. 

\emph{Side-By-Side} was found to be informative as \textit{``everything is around you''}. Some participants also liked to be able to switch the view by simply rotating their heads, while some also complained that this body movement to be \textit{``a bit tired''}. 

\emph{Toggle} was found to be ``efficient'' to switch the view, but \textit{``boring''}. Many participants found difficulties in switching to the desired view: \textit{``I eventually just slammed the button in any direction until I found the one I was looking for''}.

\vspace{0.5em}
\noindent\textbf{Key Findings:}
The main finding of the second study was that participants were generally more accurate with \emph{Tilt Map} without paying a significant cost in time. Specifically:
\begin{itemize}[leftmargin=3mm]
    \item \emph{Tilt Map} was more accurate than \emph{Side-By-Side} and faster than \emph{Toggle}.
    \item \emph{Tilt Map} was more accurate than the single \emph{Choropleth} and single \emph{Coloured Prism} for (the region task) with the small dataset (US).
    \item \emph{Side-By-Side} was faster than \emph{Toggle}.
 \end{itemize}   
We also found that \emph{Tilt Map} was strongly preferred for visual design while \emph{Toggle} was least preferred for both visual design and readability.
Generally, participants felt more confident with both \emph{Tilt Map} and \emph{Side-By-Side} than \emph{Toggle}.

Analysis of user interaction revealed that participants had different strategies of using different views in the three visualisations. 
We were surprised to find that some participants used the transition views between the main views in \emph{Tilt Map} to answer the questions.

\section{Discussion and Conclusion}

We have investigated the presentation of area-linked data in immersive visualisation environments. While data visualisation is currently rare in VR or AR environments, we believe it will become much more common with the commodification of HMD VR and AR displays~\cite{chandler2015immersive,dwyer2018immersive,Marriott2018}. 
\added{Our long term goal is not only VR but mixed reality (MR), which includes VR and AR. We foresee MR eventually supplanting 2D displays in many situations, just as current mobile devices have displaced traditional computing platforms. In particular, MR supports remote collaboration and situated data analysis, which opens up exciting new possibilities for visualisation away from the desktop.}
Our research therefore provides a timely insight into how best to visualise area-linked data in MR.

In our first study, we found that encoding quantitative data with prism height was more accurate but slower than with colour in more traditional 2D choropleth maps. 
\added{
However, it was clear that occlusion and perspective distortion still hampered understanding of the prism maps. Our results suggest that participants had to spend extra time with the prism map to find occlusion-free viewing points. While occlusion is inevitable in prism maps, the embodied interaction of changing viewing point and direction in the immersive environment appears to alleviate this issue.}

In our second study, we investigated hybrid representations that combined 2D choropleth, 3D prism map and 2D bar chart views. We compared the \emph{Tilt Map}---a new kind of interactive visualisation whose choice of idiom depends upon the user's viewing angle---with side-by-side placement and interactive toggling. We found benefits in time, accuracy and user preference for the \emph{Tilt Map} over these alternatives. 
\added{We believe, compared to our tested alternatives, the continuous embodied transition from the \emph{Tilt Map} reduces the cognitive load of context-switching between views. 
In side-by-side placement, the user has to visually link separate display spaces; and in interactive toggling, the immediate transition makes it difficult to track objects. However, further research is needed to test how effective the morphing is in the \emph{Tilt Map} for object tracking. }
\added{Due to the limitations of visual working memory, the user can only track a limited number of areas in the \emph{Tilt Map}. This potentially explains why we found that the advantage of \emph{Tilt Map} was larger in the US data (five areas to track) than in the EU data (ten areas to track).
}

Although we have not seen them used before, we believe that orientation-dependent visualisations like the \emph{Tilt Map} are widely applicable in immersive environments. Tilting provides a natural, embodied interaction for view switching that allows the choice of view to simply adapt to and take advantage of the viewing angle. 
Another example would be for an interactive map showing area-linked data. When held vertically in front of the user this could show a traditional 2D proportional symbol map. When tilted it could transition to a map with height-varying 3D bars, then transitioning like the \emph{Tilt Map} to a bar chart. This is something we plan to explore further and also test applicability to non-geographic and non-spatial data visualisation.

Limitations of our two studies are that only population data was tested and only a limited number of tasks were trialled. The available interactions were also deliberately limited. 
Future work will be needed to add standard view interactions such as filtering, selection, and zooming.
 \added{Future work should also investigate the impact of the participant's field of view. For example, in the area-comparison task of the first study, the regions in the far and close conditions in the UK map are roughly within the foveal field of view while in the far condition of the US map they may not be.}
\added{Additionally, we used a sequential colour scheme from ColorBrewer~\cite{brewer:1997}. Colour schemes are well studied for traditional cartography, but there are no clear guidelines for using them in an immersive environment, where the use of shading may interfere with color perception. 
In AR with optical see-through rendering, other issues arise, for example, colour contrast between virtual objects and the real world backdrops is often too weak.}

One limitation of the \emph{Tilt Map} and other hybrid views used in the second study was that the bar chart view does not scale well to large numbers of areas. 
\added{With the resolution of current HMD displays, we were able to visualise charts with approximately 100 bars.} 
For larger numbers of bars, one possible solution would be to split the bar chart into multiple rows if it contains more bars than can be shown within the field of view.

\added{
We would also like to explore if other ways of determining the order of the bar chart may bring extra benefits.
For example, in Fig.~\ref{fig:Exp-02-magic} there is a vertical sequence of areas where Texas is followed by Oklahoma, Kansas, Nebraska, South Dakota, and North Dakota. The bar chart currently displays the corresponding bars in an apparently random order, but it could possibly retain the vertical order of the prism map. 
Alternatively, it would also make sense to ``gridify'' the bar chart to reflect the geography, similar to spatial treemaps~\cite{wood:2008} or grid maps~\cite{mcneill_generating_2017}. Thus, if the map was oriented with north at the top, the assigned row would reflect the north-south position of an area, and the assigned column would reflect its east-west position. These alternative arrangements might simplify object tracking when the \emph{Tilt Map} transitions to the bar chart. 
}

\ifCLASSOPTIONcompsoc
  \section*{Acknowledgments}
\else
  \section*{Acknowledgment}
\fi

This work was supported by the Australian Research Council through grants DP140100077 and DP180100755.
Yalong Yang was partially supported by a Harvard Physical Sciences and Engineering Accelerator Award. We also wish to thank all our participants for their time and our reviewers for their comments and feedback.

\ifCLASSOPTIONcaptionsoff
  \newpage
\fi

\bibliographystyle{IEEEtranDOI}
\bibliography{sample}
\vspace{-4.5em}

\begin{IEEEbiography}[{\includegraphics[width=1in,height=1.25in,clip,keepaspectratio,trim=2mm 0mm 2mm 0mm]{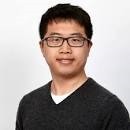}}]{Yalong Yang}
is a Postdoctoral Fellow at the John A. Paulson School of Engineering and Applied Sciences, Harvard University. He obtained his PhD from Monash University, Australia in Jan 2019 and continued as a Postdoc at Monash before joining Harvard in Oct 2019. His research designs and investigates interactive information visualisations on both conventional 2D screens and in 3D immersive environments. He received an Honorable Mention Award at InfoVis 2016. 

\end{IEEEbiography}
\vskip -3\baselineskip plus -1fil
\begin{IEEEbiography}[{\includegraphics[width=1in,height=1.25in,clip,keepaspectratio,trim=5mm 0mm 5mm 0mm]{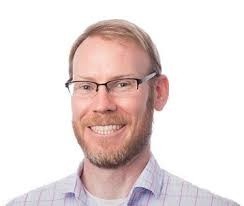}}]{Tim Dwyer}
is a Professor at Monash University, Australia, where he leads the Immersive Analytics and Data Visualisation research group.  He received his PhD from the University of Sydney in 2005, was a post-doctoral Research Fellow at Monash, then from 2008 he moved to Microsoft, Redmond, USA as a Visiting Researcher and then Senior Software Development Engineer.  He returned to Monash in 2012 as a Larkins Fellow.
\end{IEEEbiography}
\vskip -3\baselineskip plus -1fil
\begin{IEEEbiography}[{\includegraphics[width=1in,height=1.25in,clip,keepaspectratio]{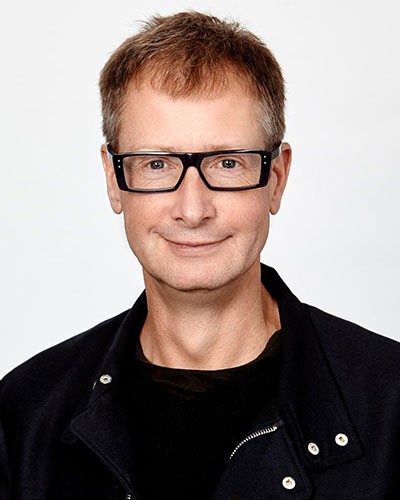}}]{Kim Marriott}
is a Professor in Computer Science at Monash University, Australia. His research is in data visualisation, human-in-the-loop analytics, assistive technologies and immersive analytics. After obtaining his PhD from the University of Melbourne in 1989, he worked at the IBM TJ Watson Research Center until joining Monash in 1993.
\end{IEEEbiography}
\vskip -3\baselineskip plus -1fil
\begin{IEEEbiography}[{\includegraphics[width=1in,height=1.25in,clip,keepaspectratio]{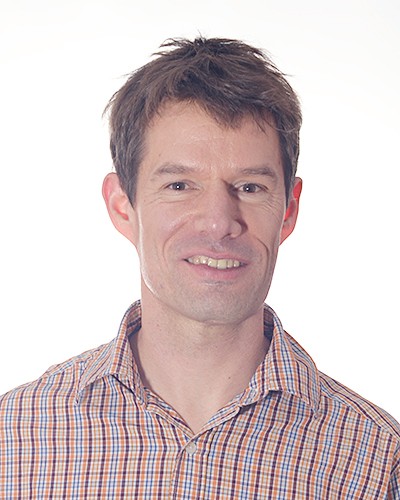}}]{Bernhard Jenny} is an Associate Professor at Monash University, Australia. He obtained a PhD in cartography from ETH Zurich, and worked in cartography and geovisualisation at Oregon State University and RMIT Melbourne. His research focuses on immersive maps for visualising and interacting with geographic data in virtual reality and augmented reality.
\end{IEEEbiography}
\vskip -2.5\baselineskip plus -1fil
\begin{IEEEbiography}[{\includegraphics[width=1in,height=1.25in,clip,keepaspectratio]{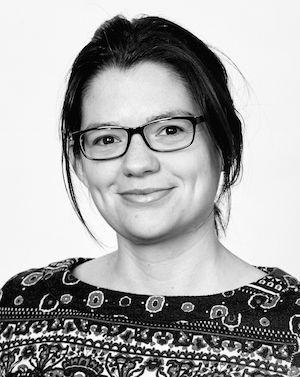}}]{Sarah Goodwin} is a Lecturer at Monash University, Australia. She
received her PhD in Geographical Information Science from City, University of London in 2015. Her research explores novel geovisualisation techniques and methodologies for visualisation design studies. Her professional and academic background is in geography, spatial analysis and geovisualisation.  
\end{IEEEbiography}

\end{document}